\documentclass[aps,prb,amsmath,twocolumn,amssymb,titlepage,superscriptaddress,shownopacs]{revtex4-2}
\usepackage[english]{babel}
\usepackage{graphicx} 
\usepackage{transparent}
\usepackage{amsmath}
\usepackage{amssymb}
\usepackage{epstopdf}
\usepackage{amsfonts}
\usepackage{bm}
\usepackage{times}
\usepackage{color}
\usepackage{mathtools}
\usepackage{hyperref}
\usepackage{tabularx}
\usepackage{hhline}
 
\newcommand{\astcycl}{\mathrlap{\kern0.085em{\circlearrowright}}\ast}
\newcommand{\taucycl}{\mathrlap{\kern0.42em{\bullet}}\circlearrowright}

\newcommand{\tcut}{t_{\text{cut}}}

\begin{document}

\title{Solving quantum impurity models in the non-equilibrium steady state with tensor trains}
\author{Martin Eckstein}
\affiliation{I. Institute for Theoretical Physics, University of Hamburg, Notkestraße 9-11, 22607 Hamburg, Germany}
\affiliation{The Hamburg Centre for Ultrafast Imaging, Hamburg, Germany}

\pacs{05.70.Ln}

\begin{abstract}
We discuss the evaluation of the integrals for intermediate-order diagrams in the self-consistent strong-coupling expansion on the Keldysh contour using Tensor Cross Interpolation (TCI). TCI is used to factorize the nested parts of the integrand, allowing the integral to be computed as a recursion of convolution integrals, which are efficiently evaluated using the Fast Fourier Transform. The evaluation of diagrams where all vertices lie on one branch of the Keldysh contour resembles the structure of the imaginary-time formalism. For diagrams with time arguments on both contour branches, we find that it can be advantageous  to parametrize the integrals in terms of physical time arguments with an additional sum over Keldysh indices. We benchmark the solution in relevant test cases, including the single impurity Anderson model and an exactly solvable electron-boson model. While the bond dimension increases with diagram order, the TCI-based integration efficiently handles low-order diagrams, making it a promising approach to go beyond the non-crossing approximation in steady-state non-equilibrium dynamical mean-field theory simulations.
\end{abstract}

\maketitle

\section{Introduction}

Solving quantum impurity problems is both a problem of intrinsic interest, and a key component in dynamical mean field theory (DMFT) \cite{Georges1996}. While equilibrium solutions can be efficiently obtained in imaginary time  using quantum Monte Carlo (QMC) methods \cite{Gull2011}, the situation in the real-time and real-frequency domain remains challenging. Numerically exact solutions of quantum impurity problems out of equilibrium include real-time quantum Monte Carlo techniques  \cite{Werner2009,Cohen2014,Cohen2015}, tensor network approaches \cite{Wolf2014}, and influence functional methods \cite{Cohen2011, Park2024, Thoenniss2023b, Thoenniss2023}. However, such approaches are often numerically costly or restricted to short times, such that real-time impurity solvers for non-equilibrium DMFT have remained primarily limited to perturbative techniques \cite{Aoki2014}.  A somewhat simpler subclass of non-equilibrium problems are non-equilibrium steady states (NESS), which are reached when a system is coupled to an environment with a bias. In a NESS, observables become time-translationally invariant, but correlation functions do not satisfy the equilibrium fluctuation-dissipation relation. This situation typically arises in transport settings, where thermal or potential gradients are applied across a quantum dot, but a NESS can also describe pre-thermal states that emerge during the transient dynamics \cite{Lange2017}. For example, photo-doped states \cite{Murakami2023} in laser excited solids \cite{Sentef2021} can be approximated as a non-equilibrium state with an effective bias \cite{Li2021}, alternative to a quasi-equilibrium approach \cite{Murakami2022}. In DMFT, this concept has been used to stabilize steady states of superconducting  and magnetic orders \cite{Li2020eta,Ray2023,Li2023}.

Also for non-equilibrium steady states in quantum impurity problems, potentially exact methods have been developed, including the auxiliary master equations approach \cite{Arrigoni2013}, weak-coupling-based QMC solvers \cite{Profumo2015, Moutenet2019}, the quasi Monte Carlo method to solve high-dimensional integrals \cite{Bertrand2021}, and the steady state variant \cite{Erpenbeck2024,Erpenbeck2023b} of the inchworm QMC algorithm \cite{Cohen2015} which was applied to simulate photo-doped Mott insulators \cite{Kunzel2024}. Nevertheless, these methods remain computationally intensive, especially when extending them to multi-orbital models. In general, a  versatile alternative for tackling quantum impurity models is the perturbative strong-coupling expansion, a systematic expansion in the coupling strength between the impurity and the bath \cite{Keiter1971, Bickers1987, Coleman1984}, which is also the basis  for a continuous time QMC \cite{Werner2006} and a bold diagrammatic QMC formulation \cite{Haule2023}. A formulation for problems out of equilibrium has been given in Ref.~\cite{Eckstein2010nca}. Most simulations for non-equilibrium DMFT remain limited to the lowest-order approximation, also called non-crossing approximation (NCA), which is restricted to certain parameter regimes and often less controlled for multi orbital problems. On the other hand, the series can be rapidly convergent, and even relatively low-order extensions of the strong coupling expansion (e.g., third-order) can provide a reasonable correction to the NCA both in equilibrium \cite{Pruschke1989, Haule2013} and for the short time evolution out of equilibrium \cite{Eckstein2010nca}. This  provides a strong motivation to efficiently evaluate  corrections beyond NCA for steady state non-equilibrium problems. 

The numerical bottleneck in this context  is the evaluation of multi-dimensional integrals, for which tensor cross interpolation (TCI)  \cite{Oseledets2011, Dolgov2020, Fernandez2022, Fernandez2024} can provide a possible solution. The TCI algorithm attempts to approximate a function of several variables by a matrix product state (MPS),
\begin{align}
F(x_1,...,x_n)=A_1(x_1)\cdots A_n(x_n),
\label{TCI}
\end{align}
with matrices  $A_i$. If the bond dimension of the MPS is low, this ansatz presents a highly compressed representation of the function. TCI constructs the MPS through successive measurements of the function at pivot points. By sweeping through the arguments $x_1,...,x_n$, new pivot points are added, and the bond dimension of the MPS is increased up to a desired accuracy.  TCI has been used for the evaluation of Feynman diagrams of the bare strong coupling expansion \cite{Erpenbeck2024} up to remarkably high orders. In this paper, we explain how TCI can be adapted for the evaluation of the diagrams in the strong coupling expansion on the Keldysh contour.  Compared to imaginary-time evaluation, two key differences emerge: (i) On the imaginary axis, the TCI integration can be directly applied to the bare expansion of the resolvent operator, similar to the QMC hybridization expansion  \cite{Werner2006}.  In real time, however, the resolvent operator decays slowly with time, and the precise form of its long time tail results from disconnected diagrams at high order. This is why we resort to the computation of the one-particle irreducible self-energy, where typically much lower orders are sufficient. (ii) Second, when using TCI to deal with functions of many variables, the rank of the decomposition depends critically on the parametrization of the function. For example, the trivial example $f(x,y)\sim\delta(x-y)$ is of rank one in the representation $\tilde f(s,y)=f(y+s,y)$, while it is full rank when decomposed  into functions depending on $x$ and $y$. In the strong coupling expansion for the steady state, the time arguments lie on the Keldysh contour rather than a single imaginary time contour, making it less clear what is  the most efficient parametrization.  Here we introduce and compare two different parametrizations to evaluate the integrals.

The paper is structured as follows: Section \ref{sec:form} provides an outline of the formalism. Details, including a brief recapitulation of the strong coupling expansion on the Keldysh contour, are given in the Appendices. Section \ref{sec:bendch} contains numerical benchmarks for the convergence of the integrals for the self-energy in a purely numerical toy model \ref{sec:convan}, the single impurity Anderson model \ref{sec:siam}, and an exactly solvable electron boson model \ref{sec:holsteinatom}. Section~\ref{sec:con} contains a conclusion and a brief outlook.

\section{Formalism}
\label{sec:form}

\subsection{Impurity model}

A general impurity model consists of a small interacting quantum system (the impurity) coupled to an infinite noninteracting environment. Examples include the Anderson model for a single orbital embedded in a fermion reservoir (Sec.~\ref{sec:siam}), or the Anderson-Holstein model which includes the coupling of a fermion to a continuum of bosonic fluctuations (Sec.~\ref{sec:holsteinatom}).  A NESS in such a system can be reached by preparing the isolated impurity in a given state at time $t=-\infty$, and thereafter switching on the coupling to the environment. Assuming that the initial transient evolution after the switch-on is damped, the system will then reside in a NESS which is independent of the initial condition and uniquely defined by the properties of the bath.

\begin{figure}[t]
\centerline{\includegraphics[width=0.4\textwidth]{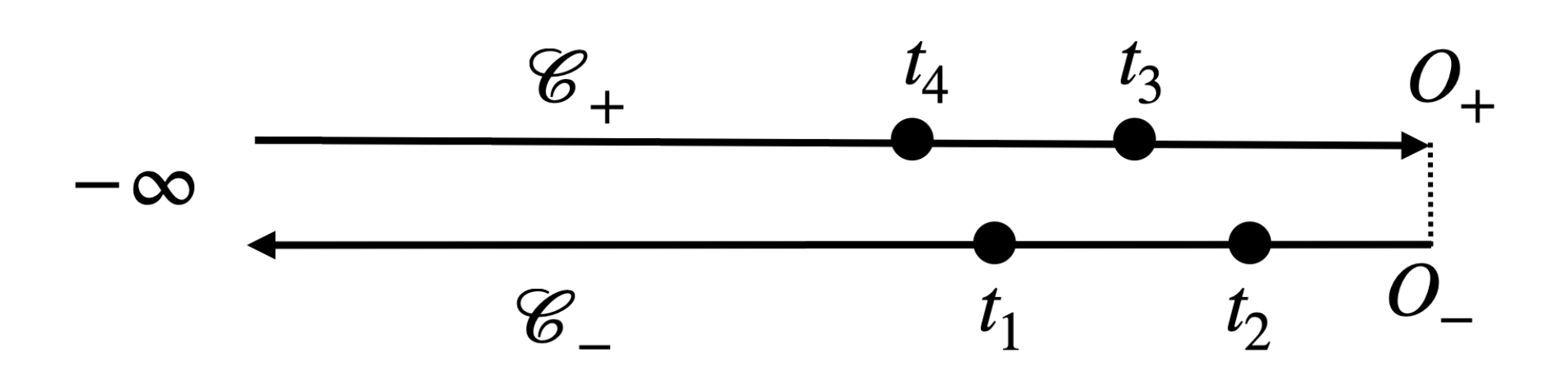}}
\caption{The two-branch Keldysh contour $\mathcal{C}$, extending from time $-\infty$ to $0$ on the upper branch $\mathcal{C}_+$ and back from $0$ to $-\infty$ on the lower branch $\mathcal{C}_-$. The contour order of the indicated time points is $t_1 > t_2 > t_3 > t_4$, the cyclic order is $t_3 \succ t_4 \succ t_1 \succ t_2$.}
\label{fig:C}
\end{figure}

To compute  properties of  this steady state without going through the transient evolution, one formulates the many-body problem using the two-branch Keldysh contour $\mathcal{C}$ \cite{KamenevBook}, extending from time $-\infty$ to $0$ on the upper branch $\mathcal{C}_+$ and back from $0$ to $-\infty$ on the lower branch $\mathcal{C}_-$ (see Fig.~\ref{fig:C}). After integrating out the environment, the impurity model is defined by an action $\mathcal{S}$, from which  expectation values of an operator $A$ at the impurity, or more general correlation functions of several operators are obtained by the contour-ordered expectation vaue
\begin{align}
\langle T_{\mathcal{C} }A(t) B(t') \cdots\rangle 
=
\frac{1}{Z}\text{tr}
\big[\rho_0
 T_{\mathcal{C}}
 e^{i\mathcal{S}}
 A(t) B(t') \cdots
\big].
\label{Cexpval}
\end{align}
Here $ T_{\mathcal{C}}$ is the contour ordering operator, which orders operators according to their time argument  with increasing times on $\mathcal{C}$, starting from $-\infty$ on $\mathcal{C}_+$ to $-\infty$ on $\mathcal{C}_-$ (see Fig.~\ref{fig:C}); $\rho_0$ is the initial state density matrix of the impurity, and $Z=\text{tr}(\rho_0 T_{\mathcal{C}}  e^{i\mathcal{S}})$  is kept as a normalization constant.

The action includes a local term $\mathcal{S}_{\text{loc}} = -\int_{\mathcal{C}} dt H_{\text{loc}}(t)$, which contains all the interactions between degrees of freedom on the impurity, and a sum of time-nonlocal interaction terms, which arise from integrating out the environment. The latter will be called ``pseudo-particle interactions'' (this terminology is explained below) and they are of the form 
\begin{align}
\mathcal{S}_{\text{int},\alpha}
=
-\int_{\mathcal{C}} d\bar t dt\,
\bar v_\alpha(\bar t) \Delta_\alpha (\bar t,t) v_\alpha(t),
\label{intgenerega}
\end{align}
with a time non-local function $\Delta$ connecting impurity operators $\bar v$ and $v$; $\alpha$ labels the type of the interaction. As the bath has been integrated out, all operators act on the local Hilbert space of the impurity site. For example, for the Anderson model in Sec.~\ref{sec:siam}, the Hilbert space has the four states $|0\rangle$, $|\!\!\uparrow\rangle$, $|\!\!\downarrow\rangle$, and $|\!\uparrow\downarrow\rangle$, and there are two interaction types $(\bar v,v) \equiv (c_{\uparrow}^\dagger, c_\uparrow)$ and $(\bar v,v) \equiv (c_{\downarrow}^\dagger, c_\downarrow)$. In a steady state problem, the bath correlation functions $\Delta_\alpha$ depend only on time difference, while both $H_{\text{loc}}$ and the  operators $\bar v$ and $v$ do not explicitly depend on time; their time argument in the action \eqref{intgenerega} just  indicates the position in the $\mathcal{C}$ ordering.

\subsection{Strong coupling expansion}
\label{sec:mainstrc}

The strong coupling expansion  \cite{Keiter1971,Bickers1987,Pruschke1989,Coleman1984} is a systematic diagrammatic expansion of impurity expectation values and correlation functions in the pseudo-particle interactions. In this section, we will only summarize the equations needed for the main focus of the article. For a more self-contained presentation, the derivation of the strong-coupling expansion on the Keldysh contour is recapitulated in App.~\eqref{app:strcpl}, mainly following Refs.~\cite{Aoki2014,Eckstein2010nca}.  In the self-consistent strong coupling expansion, all quantities are expressed in terms of ``resolvent operators'', or ``pseudo-particles propagators'' $\mathcal{G}(t,t')$ \footnote{The strong coupling expansion can also be derived by introducing pseudo-particles for each many body state $|m\rangle$ on the impurity \cite{Coleman1984}. In this formulation, the resolvents $\mathcal{G}$ have the meaning of ``pseudo-particles propagators'' and $\mathcal{S}_{\text{int}}$ corresponds to a ``pseudo-particle interaction''. We will adopt this terminology even though the derivation presented in App.~\eqref{app:strcpl} does not introduce pseudo-particles.}.  Up to pre-factors, the bare propagators $\mathcal{G}_0$ are time-evolution operators for the isolated impurity site, while $\mathcal{G}$ is a dressed time-evolution operator between times $t$ and $t'$. All pseudo-particle propagators are matrices with the dimension of the local impurity Hilbert space. \footnote{Symmetries imply a block-diagonal structure of these matrices, which is not indicated in the equations below, but used in the numerical implementation}. 

The pseudo-particle propagators are determined by two self-consistent equations: (i) The expression of a pseudo-particle self-energy $\Sigma(t,t')$ in terms of $\mathcal{G}$ and $\Delta$, and (ii), the Dyson equation; $\Sigma$ represents a correction to the time evolution, where the state on the impurity successively undergoes transitions induced by the interaction with the environment, and the Dyson equation incorporates an arbitrary number of these corrections into the resolvents $\mathcal{G}$. The Dyson equation is an integral equation of the form
\begin{align}
\mathcal{G}&(t_4,t_1)=\mathcal{G}_0(t_4,t_1)\,\,+
\nonumber
\\
&+
\int_{t_4\succ t_3\succ t_2\succ t_1} 
\!\!\!\!\!\!\!\! \!\!\!\!\!\!\!\!\!\!
dt_3 dt_2\, \,\,\,\mathcal{G}_0(t_4,t_3) \Sigma( t_3,t_2)\mathcal{G}(t_2,t_1),
\label{dyson}
\end{align}
which must be solved with the initial condition  $\mathcal{G}(t,t')=-i \text{~~for ~~}t\to t'$. Because the strong coupling expansion is an expansion of the time-evolution operator, the times in the integrals have a definite order: Following Ref.~\cite{Eckstein2010nca}, we adopt the so-called cyclic ordering, which starts from time $0_-$ on $\mathcal{C}_-$, proceeds to  to $-\infty$ on $\mathcal{C}_-$, and back from $-\infty$ to $0_+$ on $\mathcal{C}_+$ (see Fig.~\ref{fig:C}); we will use the notation $t\succ t'$ to indicate that $t$ is later than $t'$ according to cyclic ordering, and the notation 
\begin{align}
\int_{\bm t_\succ} d\bm t 
=
\int_{t_n\succ ...\succ t_1}  dt_2 \cdots t_{n-1},
\,\,\,\,\bm t=(t_n,...,t_1),
\end{align}
for the integration over variables $dt_2 \cdots t_{n-1}$ respecting the cyclic ordering $t_n\succ ...\succ t_1$, such as in Eq.~\eqref{dyson}.

\begin{figure}[tbp]
\centerline{\includegraphics[width=0.4\textwidth]{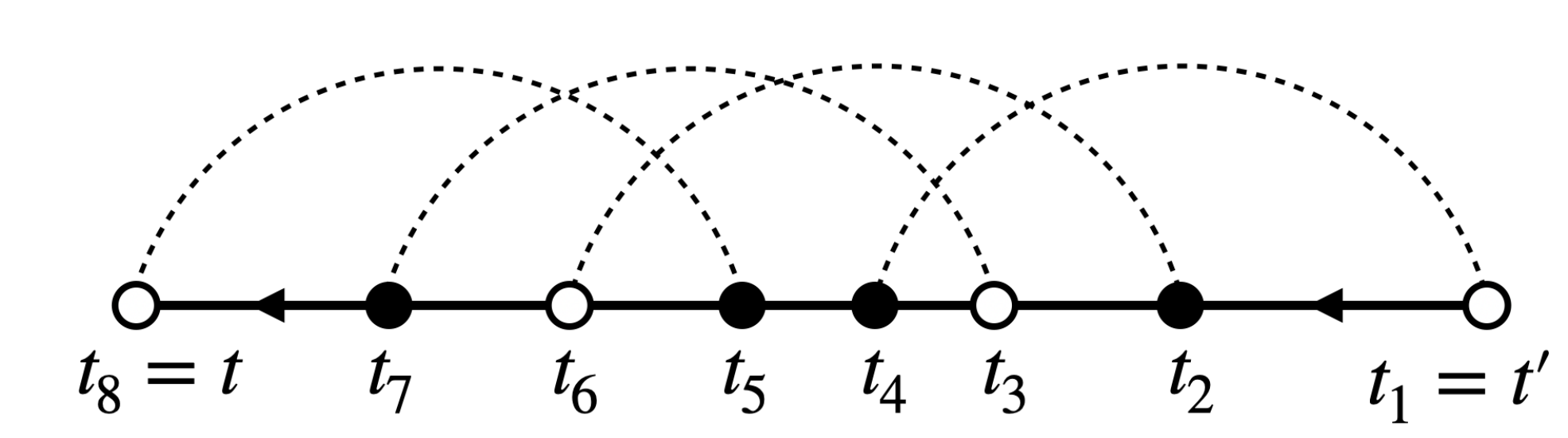}}
\caption{a) A diagram for the  self-energy $\Sigma(t,t')$ ($t \succ t'$) of order $m=4$.
Solid lines represent propagators $\mathcal{G}(t_{j+1},t_j)$ between successive time points, empty (filled) dots represent operators $v$ ($\bar v$), and a  dashed line directing from an open dot at $t_i$ to a closed dot at $t_j$ represents an interaction line $\Delta (t_j,t_i)$ .
 }
\label{Sdiagram}
\end{figure}

The self-energy $\Sigma(t,t')$ can be computed as a sum over contributions from different diagrams $\mathcal{D}$. A self-energy diagram of order $m$ has the form as indicated in Fig.~\eqref{Sdiagram}: The back bone represents propagators $\mathcal{G}$ (solid lines) between time points $t_1,...,t_n$, separated by vertices (interaction events) $v_i^{\mathcal{D}}$ and $\bar v_j^{\mathcal{D}}$ (open and filled dots, respectively). In addition there are $m$ interaction-lines (dotted lines), which connect the interaction events. The self-energy diagram has the analytical expression 
\begin{align}
\Sigma_{\mathcal{D}}
(t_n,t_1) 
&= 
\int_{\bm t_\succ} d\bm t 
 \,\,\,F_{\mathcal{D}}(\bm t)
 W_{\mathcal{D}}(\bm t),
 \,\,\,\,\bm t=(t_n,...,t_1),
  \label{SDegedws01}
\end{align}
where integration variables have a cyclic order,
\begin{align}
 W_{\mathcal{D}}(\bm t)
 = v^{\mathcal{D}}_{n} \mathcal{G}(t_n,t_{n-1})v^{\mathcal{D}}_{n-1} \cdots v^{\mathcal{D}}_2 \mathcal{G}(t_2,t_{1}) v^{\mathcal{D}}_1
\label{wfunc}
\end{align}
is the product of successive propagators and vertices, and 
\begin{align}
\label{ffunc}
F_{\mathcal{D}}(\bm t)
=
C_{\mathcal{D}} \prod_{j=1}^m \Delta_{\alpha_j}(t_{\bar a_j},t_{a_j})
\end{align}
combines the product of interaction lines and an overall pre-factor $C_{\mathcal{D}} $, assuming that the labelling is such that  line $j$ directs from the vertex $v_{\alpha_j}^{\mathcal{D}}$ at time $t_{a_{j}}$ to the vertex $\bar  v_{\alpha_j}^{\mathcal{D}}$ at time $t_{\bar a_{j}}$. The diagrams for the self-energy are both one-particle irreducible (they cannot be separated by cutting a single $\mathcal{G}$ line)  and skeleton-like, i.e., $\mathcal{G}$ lines in the diagram do not have self energy insertion themselves. Inserting the self-energy self-consistently inside the Dyson equation then generates all terms of the bare expansion. The first order diagrams, which define the so-called non-crossing approximation (NCA), contains no internal integration.  The derivation of these equations, as well as the rules to generate all diagrams at a given order, are reviewed in App.~\ref{app:strcpl}. 

The main numerical challenge in the evaluation of the self-consistent strong expansion  is the solution of the high-order integrals \eqref{SDegedws01} for the self-energy. There are $2m-2$ integrals to compute $\Sigma$ for each external time argument, leading to an overall scaling $\mathcal{O}(N^{2m-1})$ of the numerical effort with the number of time points $N$ if integration is based on an equidistant grid with $N$ time points of $\mathcal{C}$. This can become numerically costly already at second order, and the main goal of this work is to evaluate these integrals more efficiently.

In addition to the evaluation of $\Sigma$, one must solve the Dyson equation \eqref{dyson}. The solution of the Dyson equation in the steady state has been discussed in Ref.~\cite{Li2021}, and we follow a similar approach here (although most of the benchmarks are focused on the evaluation of the high-order integrals and are therefore done with a fixed input instead of within self-consistent calculation). Reference \cite{Li2021} uses a solution in frequency space, with a numerical effort $\mathcal{O}(N\log N)$ due to the use of fast Fourier transforms (FFT). In App.~\ref{App:dyson} we describe a solution in real time, which at the same numerical cost $\mathcal{O}(N\log N)$  allows to incorporate the quadrature rules of higher order accuracy which are used for the solution of the Dyson equation in the two-time formalism within the NESSi library \cite{Nessi}.

\subsection{Lesser and greater components of $\mathcal{G}$}

Before going into the description of the self-energy, let us define the parametrization of contour functions in terms of functions depending on physical time arguments. For the bath correlation functions $\Delta$ we use the standard parametrization $\Delta(t,t') =  \Delta^>(t,t')$ if $t$ is later then $t'$ according to the ordering $T_\mathcal{C}$, and $\Delta(t,t') \equiv  \Delta^<(t,t')$ otherwise. In a steady state these functions depend on physical time arguments only. For pseudo-particle propagators we will use a similar terminology,
\begin{align}
\mathcal{G}(t,t')
&=\begin{cases}
\mathcal{G}^>(t-t')
&\text{~~for~~}
t,t' \in\mathcal{C}_-, t<t'
\\
&\text{~~and~~}
t,t' \in\mathcal{C}_+, t>t'
\\
\mathcal{G}^<(t-t')
&\text{~~for~~}
t \in\mathcal{C}_+, t'\in\mathcal{C}_-
\end{cases},
\label{lesgtrparam}
\end{align}
and analogous for $\Sigma$ and $\mathcal{G}_0$. Other combinations (such as $t,t' \in\mathcal{C}_+$ with $t<t'$) do not appear in  Eqs.~\eqref{dyson} and \eqref{SDegedws01}, due to the cyclic ordering. Moreover, we note that propagators  are defined such that they satisfy a hermitian property
\begin{align}
X^>(t) = - [X^>(-t)]^\dagger,
\,\,\,\,
X^<(t) = - [X^<(-t)]^\dagger,
\label{lesgtrparamherm}
\end{align}
which relates  $X=\mathcal{G},\Sigma,\mathcal{G}_0$ with positive and negative time arguments.

Due to the time-ordered structure of the integrals in Eqs.~\eqref{dyson} and \eqref{SDegedws01}, one can see that the equations for the greater components  of the pseudo-particle propagators are decoupled from the lesser ones:  We can choose a time  $t<0$ on $\mathcal{C}_-$, such that $\Sigma^>(t)=\Sigma(t_{-},0_-)$, $\mathcal{G}^>(t)=\mathcal{G}^>(t_{-},0_-)$. With this, $\Sigma^>(t)$ and $\mathcal{G}^>(t)$  depend  through Eqs.~\eqref{dyson} and \eqref{SDegedws01} only on $\mathcal{G}^>(\bar t)$ and $\Sigma^>(\bar t)$ with $\bar t \in  [t,0]$. We can therefore first compute an independent self-consistent solution for $\Sigma^>$ and $\mathcal{G}^>$. This solution then enters the computation for $\Sigma^<$ and $\mathcal{G}^<$, and  both $\mathcal{G}^>(t)$ and $\mathcal{G}^<(t)$ are then needed to compute physical correlation functions and expectation values.

\subsection{Evaluation of $\Sigma^>$}

Here we describe the computation of $\Sigma^>(t)$ at a time $t<0$, which is obtained from the integral \eqref{SDegedws01} with external time arguments $t$ on $\mathcal{C}_-$ and $t'=0$ on $\mathcal{C}_-$. The factor $W_{\mathcal{D}}(\bm t)$ in the integrand is a product of functions which depend only on the time differences between successive time arguments.  We therefore write the functions $F$ and $W$ in the integrand in Eq.~\eqref{SDegedws01} in terms of these time differences $\tau_j=t_j-t_{j+1}$, $j=1,...,n-1$. For convenience we choose $\tau_j$ positive, such that $\sum_{l}\tau_l=-t$, and for simplicity use the notation $F_{\mathcal{D}}(\bm \tau)\equiv F_{\mathcal{D}}(\bm t(\bm \tau))$ to denote the function $F$ written in terms of the variables $\bm \tau$ (similar for $W$). While $W_{\mathcal{D}}(\bm \tau) $ factorizes into functions depending on only one argument  $\tau_j$, the function $F_{\mathcal{D}}(\bm \tau)$ is typically fully nested. At this point, we can use the TCI to attempt a matrix product factorization in the form \eqref{TCI},
\begin{align}
\label{ffac}
F_{\mathcal{D}}(\bm \tau)=A_{n-1}(\tau_{n-1})\cdots A_{1}(\tau_{1}),
\end{align}
with matrices $A_j$. With this, the whole integrand has a factorized form, 
\begin{align}
\label{ifac}
F_{\mathcal{D}}(\bm \tau)W_{\mathcal{D}}(\bm \tau)=B_{n-1}(\tau_{n-1})\cdots B_{1}(\tau_{1}),
\end{align}
where the matrices $B_j$ are tensor products
\begin{align}
B_{j}(\tau_j)= 
\begin{cases}
[\mathcal{G}^>(-\tau_j)v^{\mathcal{D}}_{j}] \otimes A_j(\tau_j) 
& j<n-1
\\
[v_{n}^{\mathcal{D}}\mathcal{G}^>(-\tau_{j})v^{\mathcal{D}}_{j}] \otimes A_j(\tau_{j}) & j=n-1 
\end{cases}.
\label{tensorB}
\end{align}
We then transform the integral $\int_{\bm t_\succ} d\bm t $ into an integral over time differences,
\begin{align}
\Sigma^>_{\mathcal{D}}(t) = \int_{\sum_{l}\tau_l=-t}  \!\!\! d\bm \tau\,\,
B_{n-1}(\tau_{n-1})\cdots B_{1}(\tau_{1}),
\label{nesterret}
\end{align}
where we used a  short notation for the integration is over $\tau_j\in[0,\infty)$ with the constraint $\sum_{l}\tau_l=-t$:
\begin{align}
\int_{\sum_{l}\tau_l=-t}  \!\!\!d\bm \tau \equiv
\int_0^ {-t}
\!\!d\tau_1 \int_0^{-t-\tau_1}\!\!\!\!\! \!\!\!\!\!d\tau_2 \cdots \int_0^{-t-\sum_{l=1}^{n-3} \tau_l}\!\!\!\!\!\!\!\!\!\!\!d\tau_{n-2},
\nonumber
\end{align}
with $\tau_{n-1}=-t-\sum_{l=1}^{n-2}\tau_l$ in the integrand on the right hand side. The nested integral \eqref{nesterret}  can be solved by a recursion
\begin{align}
&C_1=B_1, 
\label{retrec01}
\\
&C_j=B_j\ast C_{j-1} \text{~for~}j >1,
\label{retrec02}
\\
&\Sigma^>_{\mathcal{D}}(t)=C_{n-1}(-t),
\label{retrec02}
\end{align}
with the retarded convolution 
\begin{align}
\label{retconv5}
(f\ast g)(\tau) = \int_0^{\tau} d\bar t f(\tau-\bar \tau) g(\bar \tau).
\end{align}
Each integral in Eq.~\eqref{retrec02} can be computed efficiently using FFT with an effort $\mathcal{O}(N\log(N))$, inserting the integration weights for high order quadrature rules at effort $\mathcal{O}(N)$ (see Sec.~\ref{FFT1}). The numerical implementation of the TCI factorization used an early version of the tensor library provided in Ref.~\cite{Fernandez2024}.

This way to compute $\Sigma^>$ is closely related  to  the determination of the imaginary time propagators in Ref.~\cite{Erpenbeck2023b}, where  one must compute a time-ordered integral similar to Eq.~\eqref{SDegedws01}  with integrals restricted to the imaginary time interval $[0,-i\beta]$ instead of the real axis. The main differences of the real and imaginary time formalism have been summarized in the introduction. In addition, there are two more technical differences worth mentioning: (i) Ref.~\cite{Erpenbeck2023b} uses TCI to directly factorize the full integrand into the form \eqref{ifac}, while he we factor only the part $F$ related to the interaction lines. The bond dimension in the two factorizations should however not differ much: Both are related by the tensor product \eqref{tensorB}, and the dimension $d_G$ of the $\mathcal{G}$ matrices is typically low. (ii) In Ref.~\cite{Erpenbeck2023b}, the time-ordered integration domain for a given external time $t$ is mapped to a rectangular domain using a  suitable mapping function, after which each integral can be computed independently with an effort of $\mathcal{O}(N)$. On the real axis, we  do not introduce the mapping function, and can  evaluate the integrals for all times up to a maximum time $\tcut =(N-1)h$ at a cost $\mathcal{O}(N\log(N))$. There is however one caveat: To sample the function $F(\bm \tau)$, the latter must be known in the full domain $\{\bm \tau: \tau_j\in [0,\tcut]\text{~for~}j=1,\ldots,n-1\}$.  For the evaluation of the self-energy  at order $m$ up to time $\tcut$, $\Delta$ must therefore be known on a larger interval  $m\tcut$. However, this is no problem in general, because anyway $\tcut$  must be chosen large enough to include the main support of $\Delta(t)$, such that $\Delta(t)$ can be set to zero outside this interval, or extrapolated by simple means. (In the examples below,  $\Delta$ is given explicitly for all times). 

\subsection{Evaluation of $\Sigma^<$ in cyclic parametrization}
\label{sec:celp}

The lesser component  for $t<0$ is computed from the integral \eqref{SDegedws01} with external time arguments $t$ on $\mathcal{C}_+$ and $t'=0$ on $\mathcal{C}_-$. This makes its evaluation rather different from $\Sigma^>(t)$: Firstly, according to the cyclic order, the integration over internal variables now extends to times $-\infty$. We thus introduce a sufficiently large negative cutoff $-\tcut$  in the integrals, which can be treated as a convergence parameter. More importantly, the  vertices can now lie on both $\mathcal{C}_+$ and $\mathcal{C}_-$, which makes it less straightforward for find a parametrization in which the integrand has a low rank factorization. One possible approach would be to adopt a decomposition of the integrand as a function of time differences in the cyclic ordering, 
\begin{align}
\tau_j =
\begin{cases}
t_{j}-t_{j+1} & \text{~~for~~}t_{j+1}\in\mathcal{C}_-
\\
t_{j+1}-t_{j} & \text{~~for~~}t_{j+1}\in\mathcal{C}_+
\end{cases}.
\label{tucycvle}
\end{align}
Here $\tau_j\in(0,\tcut)$ if both $t_{j}$ and $t_{j+1}$ are on the same contour, and $\tau_j\in(-\tcut,\tcut)$ for the single segment for which  $t_{j+1}\in\mathcal{C}_+$ and $t_{j}\in\mathcal{C}_-$. With this approach, one can attempt a factorization analogous to Eqs.~\eqref{ffac} and \eqref{ifac}. We will refer to this as the {\em cyclic parametrization}.  Also in this parametrization, the integral corresponding to \eqref{nesterret} can be transformed into a recursive set of convolutions. In short, if  we denote by $p$  the last argument on the lower contour,  one can recursively compute  $B_{p-1} \ast \cdots \ast B_2 \ast B_1$ from the right, and $B_{n-1}\ast B_{n-2} \ast \cdots \ast B_{p}$ from the left, and then contract the result. All integrals can again straightforwardly  be cast into convolutions which can be computed using FFT. Because the cyclic parametrization will  turn out to be less favorable for the benchmark examples, these equations will not be presented here.

\begin{figure}[tbp]
\centerline{\includegraphics[width=0.4\textwidth]{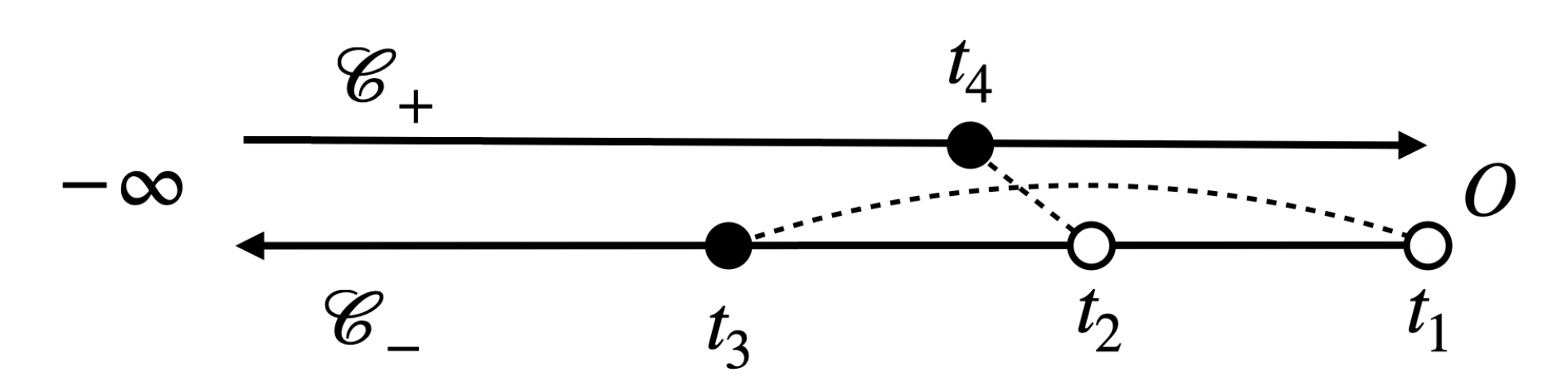}}
\caption{A contribution to $\Sigma^<(t_4,t_1)$; see text for discussion.}
\label{fig:leproblem}
\end{figure}

As it will turn out in the examples below, the matrix dimension needed to represent the function $F(\bm \tau )$ for the lesser component in cyclic parametrization can be much higher than for the greater component. This can be understood, because  the parametrization in time differences \eqref{tucycvle} is not well suited to resolve strongly time-localized features in the interactions $\Delta$.  To illustrate this fact, we consider the contribution to the $2$nd order diagram shown in Fig.~\ref{fig:leproblem}, and assume the interaction line has a contribution which is strongly localized in time on top of  a smooth background $a(t)$. As an extreme case, assume a form $\Delta^>(t)= a(t) + b\delta(t)$. The  function $F(\bm t)=\Delta^>(t_3-t_1)\Delta^>(t_4-t_2)$ then has a contribution $F_s(\bm t)= ba(t_3)\delta(t_4-t_2)$, which in cyclic  parametrization \eqref{tucycvle} becomes
\begin{align}
F_s(\tau_3,\tau_2,\tau_1) 
&= b \,a(-\tau_1-\tau_2)\delta(\tau_3-\tau_2).
\end{align}
Assuming a factorization \eqref{ffac}, we can integrate over $\tau_1$ on both sides of the equation, leading to
\begin{align}
 b \bar a(\tau_2)\delta(\tau_3-\tau_2) \stackrel{!}{=} A_3(\tau_3)\bar A_2(\tau_2),
 \label{sgrejs}
\end{align}
with $\bar a(\tau_2)=\int_0^\infty d\tau_1 a(-\tau_1-\tau_2)$ and  $\bar A_2(\tau_2)= A_2(\tau_2)\int _0^\infty d\tau_1A_1(\tau_1) $. The left hand side of \eqref{sgrejs} is a diagonal matrix, whose rank is  given by the number of non-vanishing matrix elements $\bar a(\tau_2)$, i.e., the temporal extent of the background $a(t)$. We can therefore  expect the matrix dimension in the factorization \eqref{ffac} to grow with the size of the time window $\tcut$. Instead, if all arguments $t_1$,...,$t_4$ are on the lower contour, as for the calculation of $\Sigma^>$, the function $F$ can be factorized with a fixed low bond dimension  for all times, depending on the complexity of the background $a(\tau)$.  From this simple argument, one can expect that a use of the cyclic parametrization for the factorization of the integrals \eqref{SDegedws01} is not efficient for large times, which is indeed demonstrated for some examples in Sec.~\eqref{sec:convan}.

\subsection{Evaluation of $\Sigma^<$ in Keldysh parametrization}
\label{sec:kelp}
 
\begin{figure}[tbp]
\centerline{\includegraphics[width=0.4\textwidth]{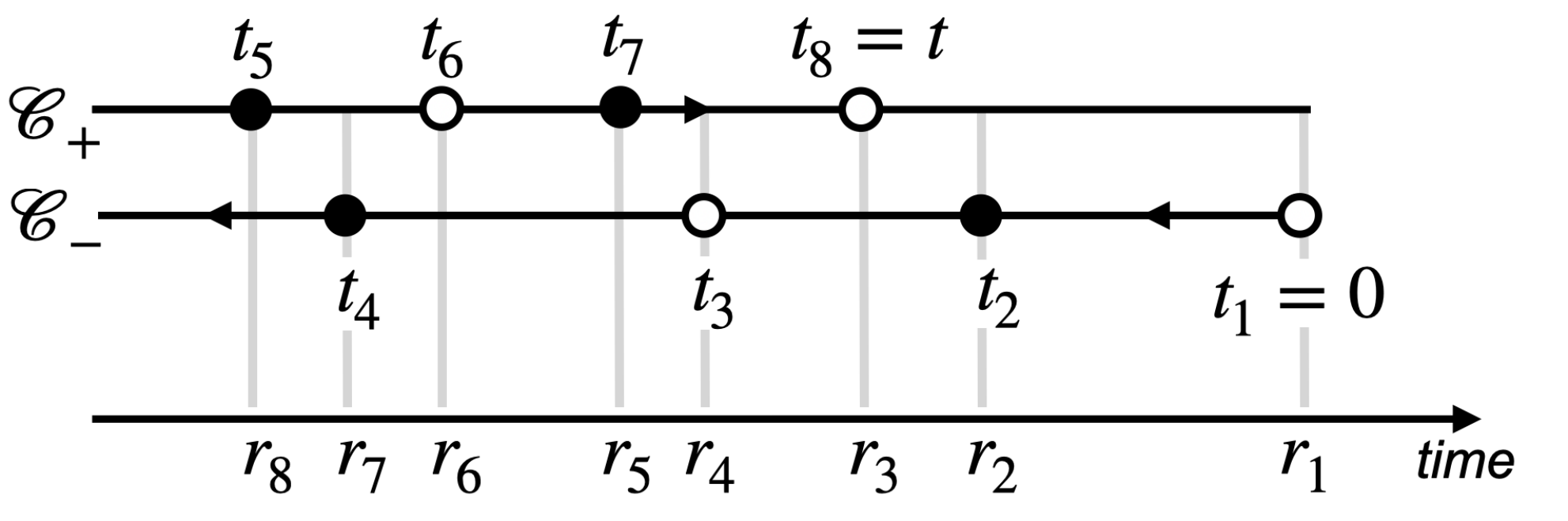}}
\caption{Reparametrization of a cyclically ordered time sequence $t_8\succ\cdots\succ t_1$ in terms of physically oredred times $r_1>\cdots>r_{8}$.}
\label{fig:keldysh_param}
\end{figure}

The unfavorable behavior of the cyclic parametrization can be overcome by a different parametrization, which, motivated by the discussion in the previous section, is based on physical time differences instead of contour time differences. Consider a time sequence as shown in Fig.~\ref{fig:keldysh_param}, with cyclic time arguments $t_n \succ \cdots \succ t_1$. We associate with it ordered physical times $0>r_1>\cdots>r_n>-\tcut$, with a given set of Keldysh indices  $(\sigma_1,...,\sigma_n)$, $\sigma_j\in \pm1$, such that vertex at time $r_j$ is on the contour $\mathcal{C}_{\sigma_j}$. For the time sequence in Fig.~\ref{fig:keldysh_param}, we have $\bm \sigma=(-,-,+,-,+,+,-,+)$.  There is a unique mapping  $(\bm r,\bm \sigma)\to\bm t$, and we denote by $s(\bm\sigma)$ the index of the external time argument, i.e., $r_{s(\bm \sigma)}\leftrightarrow t_n$ ($s=3$ in the example above). The integral \eqref{SDegedws01} can then be rewritten as 
\begin{align}
\sum_{\bm \sigma}{}' \int_{\bm r | r_{s(\bm\sigma)}}  d \bm r 
F_{\bm \sigma}(\bm r) 
W_{\bm \sigma}(\bm r) 
= \int_{\bm t _{\succ}} d\bm t 
F(\bm t)
W(\bm t),
\label{ksum}
\end{align}
where $\sum_{\bm \sigma}'$ is the sum over all sets of Keldysh indices with $\sigma_1=-1$, and
\begin{align}
\int_{\bm r | r_s} d\bm r
=
&(-1)^{s-2}\int _{-\tcut}^{r_s} dr_{n} \int _{r_{n}}^{r_s} dr_{n-1} \cdots  \int _{r_{s+2}}^{r_s}  dr_{s+1}
\nonumber\\
&\times\,\,\int _{r_s}^{0} dr_{s-1} \int _{r_{s-1}}^{0} dr_{s-2} \cdots  \int _{r_{3}}^{0}  dr_{2} 
\label{hacfadxda}
\end{align}
is the integral over physically ordered times, where $r_s$ is the external time argument, and the factor $(-1)^{s-2}$  is due to the integration directions on $\mathcal{C}_-$. After this, we  transform to physical time differences $s_{j}=r_{j}-r_{j+1}\in[0,\tcut]$, and use TCI to factorize the integrand 
$I_{\bm \sigma}(\bm r) \equiv F(\bm t(\bm r,\bm \sigma)) W(\bm t(\bm r,\bm \sigma))$,
\begin{align}
I_{\bm \sigma}(\bm r)  = B^{\bm\sigma}_{n-1}(r_{n-1}-r_n)\cdots B^{\bm\sigma}_{1}(r_1-r_2).
\end{align}
The integral \eqref{hacfadxda} can then be evaluated by a recursion with retarded convolutions \eqref{retconv5} on $[0,\tcut]$,
\begin{align}
&L_1^{\bm\sigma}=B^{\bm\sigma}_1,\,\,\, 
L_j^{\bm\sigma}=B^{\bm\sigma}_j\ast L_{j-1},\,\,\, j>1
\\
&R^{\bm\sigma}_{n}=1,\,\,\,
R^{\bm\sigma}_{j}=R_{j+1}\ast B^{\bm\sigma}_{j}, \,\,\,j<n
\\
&
\int_{\bm r | r_s} 
\!\!\!\!
d\bm r I_{\bm \sigma}(\bm r) 
=
(-1)^{s-2}
R^{\bm\sigma}_{s}(\tcut+r_{s})
L^{\bm\sigma}_{s-1}(-r_{s}).
\end{align}
If $I_{\bm \sigma}(\bm r) $ is a matrix-valued function, we can use two more tensors 
 \begin{align}
 I_{\bm \sigma}(\bm r)_{a,b} 
 =
 B^{\bm\sigma}_{n}(a)  B_{n-1}^{\bm\sigma}(r_{n-1}-r_{n})\cdots B^{\bm\sigma}_{1}(r_1-r_2) B^{\bm\sigma}_{0}(b).
 \nonumber
\end{align}
The remaining steps of the algorithm are adapted in a straightforward manner. 

\subsection{Expectation values and correlation functions}
\label{sec:Cmain}

Physical expectation values and correlation functions can be computed once a self-consistent solution for the propagators $\mathcal{G}$ has been reached.  The expectation value of an operator $A$ is obtained by inserting $A$ at time $0$ in the contour-ordered expectation value \eqref{Cexpval}, and expanding the action. This gives
\begin{align}
\langle A \rangle=  
\frac{i}{Z}
\text{tr}[
A \xi \mathcal{G}^<(0)]
\label{pseudoexp}
\end{align}
where $\xi$ is the Fermion parity operator, and the normalization is given by
\begin{align}
Z
=
i\text{tr}[
\xi \mathcal{G}^<(0)].
\label{norm}
\end{align}
 A more detailed explanation, including the origin of the fermion parity operator, is given in App.~\ref{app:strcpC}. 

The diagrams for the  correlation functions
\begin{align}
C_{AB}(t,t')=-i\langle T_{\mathcal{C} }A(t) B(t') \rangle
\label{Cab}
\end{align}
have additional vertices $A$ and $B$ on the contour. The resulting diagrams are closely related to a self-energy diagram with an additional line ``$\Delta=1$'' connecting the external vertices $A$ at time $t$ and and $B$ at time $t'$, because the correlation function can be obtained as the derivative of $Z$ with respect to the function $c(t,t')$ in a source term $\mathcal{S}_{c} = -\int_{\mathcal{C}} d\bar t dt\, A(\bar t) c (\bar t,t) B(t) $ within the action. For that reason, the evaluation of $C$-diagrams is very similar to the evaluation of diagrams for $\Sigma^<$, and will be performed in the Keldysh parametrization. Detailed expressions are given in App.~\ref{app:strcpC}. 

In the results below, we will refer to an $m$th order diagram for $C$ as one which includes $m$ lines including the virtual line between $A$ and $B$, i.e., $m-1$ interaction lines. The leading contribution to $C_{AB}(t,t')$ (NCA contribution), without any additional interaction lines, is therefore $m=1$ [Eq.~\eqref{cnca}].

\section{Benchmarks}
\label{sec:bendch}

\begin{figure}[tbp]
\centerline{
\includegraphics[width=0.5\textwidth]{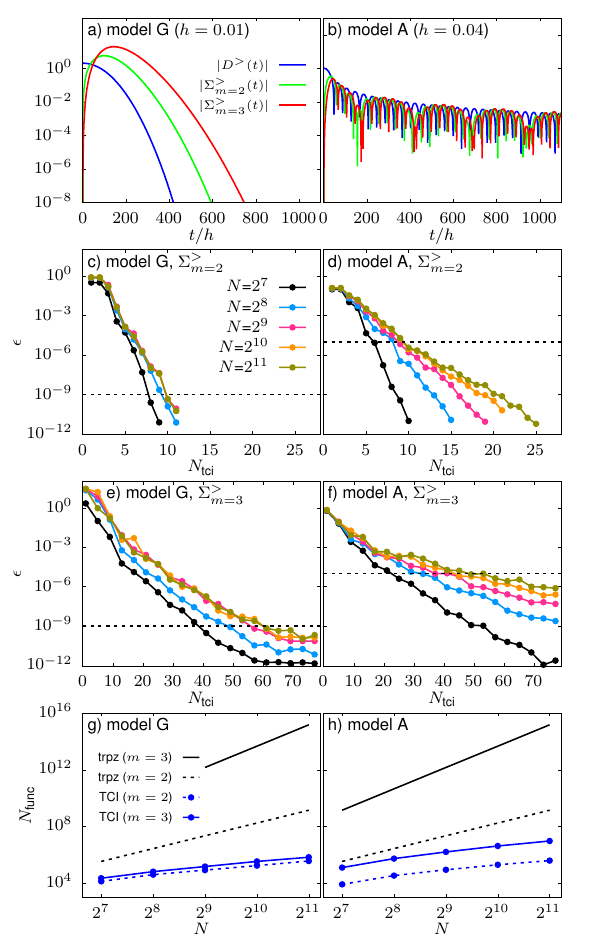}
}
\caption{
a) and b) Interaction $\Delta(t)$ and the resulting self-energies at order $m=2$ and $m=3$ for model G ($\Delta$ with Gaussian decay) and model A ($\Delta$ with algebraic decay); see legend in panel a). c)-f) Convergence of the integrals for $\Sigma^>(t)$  at diagram order $m=2$ (upper row) and $m=3$ (lower row) with the number of TCI sweeps $N_{\rm tci}$, for model G (left) and model A (right). The different curves correspond to the evaluation of the  integral on time intervals of different length, where $N$ is the number of mesh points at fixed time-step $h$; see legend in c). g) and h) Number of function evaluations in the TCI search needed to suppress the error below a given threshold ($\epsilon=10^{-9}$ for model G, and  $\epsilon=10^{-5}$ for model A, respectively, shown by the horizontal dashed lined in c) to f)). The black lines (labelled trpz) show the number of function evaluations for a direct integration.}
\label{fig:tci_convergence}
\end{figure}

\subsection{Convergence analysis}

\label{sec:convan}

To demonstrate some aspects of the algorithm, we first analyze the convergence of the integrals for $\Sigma^>$ and $\Sigma^<$ for a simple test case where $\mathcal{G}$ and $\Delta$ are given scalar functions, and all vertices  $v$ and $\bar v$ are set to unity.  (In other words, diagram topologies correspond to an impurity model with a trivial Hilbert space of only one state and an action $\mathcal{S}_{\rm int}=-\frac{1}{2}\int dtdt'\, \hat I(t)  \Delta(t,t') \hat I (t')$  with the identity operator $\hat I$. We analyze two models, in which the hybridization functions have a Gaussian decay (``model G'') and an algebraic decay (``model A''), respectively. Specifically, for model G we take  $\Delta^>(t)=-i e^{-t^2}(e^{i0.28 t}+e^{i0.23 t})$, $\Delta^<(t)=-i e^{-t^2}(e^{i0.2 t}+e^{-i0.44 t})$, $\mathcal{G}^>(t)=-i e^{-0.01t^2}(e^{i t}+e^{i0.33 t})$, and  $\mathcal{G}^<(t)=-i e^{-0.01t^2}(e^{-0.14i t}+e^{i0.2t})$, and for model A we take   $\Delta^>(t)=-i J_1(2t)/t$, $\Delta^<(t)=\Delta^>(t)e^{-i0.33 t}$, $\mathcal{G}^>(t)=-i(e^{-i0.14 t}+e^{-i0.74 t})/2$, and $\mathcal{G}^<(t)=-i(e^{-i0.3 t}+e^{-i0.4 t})/2$. The input functions and the resulting self-energies of $2$nd and $3$rd order are shown in Fig.~\ref{fig:tci_convergence}a) and b). For the simple test problems analyzed here, we also note that the third order is larger than the second order. However, a convergence of the full strong coupling series with order $m$  is not important for the purpose of the following convergence analysis at fixed order.

 In Fig.~\ref{fig:tci_convergence}c)-f) we show the convergence of the integrals for $\Sigma^>$ with the number $N_{\rm tci}$ of TCI iterations, which is essentially identical to the maximum bond dimension in the resulting matrix product state. We also compare simulations for different length $\tcut$ of the time interval, i.e., different numbers $N$ of time mesh-points at given time-step $h=0.01$.  In each curve, the result for the largest $N_{\rm tci}^{\rm max}$ is taken as a reference, and the error for smaller $N_{\rm tci}$ is evaluated as the norm
 \begin{align}
\label{norm2}
 \epsilon(N_{\rm tci}) 
 &= 
 ||
 X_{N_{\rm tci}}-X_{N_{\rm tci}^{\rm max}}
 ||_0
 \\
 &\equiv
\text{max}_{j=0}^{N-1} \big| X_{N_{\rm tci}}(jh)-X_{N_{\rm tci}^{\rm max}}(jh)\big|
 \end{align}
 over all time grid-points for the respective quantity $X$. 

First one can observe that in both model  G (left column in Fig.~\ref{fig:tci_convergence}) and A (right column in Fig.~\ref{fig:tci_convergence}), the convergence at the same order is faster for order $m=2$ than for order $m=3$, which is expected because of the larger complexity of the functions to be decomposed. However, in both cases one reaches an exponential decrease of the error with $N_{\rm tci}$. For the minimum bond dimension  $\chi_{N,\epsilon}$ which is needed to bring the error below a given threshold $\epsilon$. For model G one observes a saturation of  $\chi_{N,\epsilon}$ for large $N$, which demonstrates that TCI can automatically detect that the relevant contribution to the integral comes from a finite time range that is determined by the exponential decay of $\Delta$, independent of $N$. (Compare the  horizontal dashed line indicating an error threshold $\epsilon=10^{-9}$ in Figs.~\ref{fig:tci_convergence}c) and e), which shows that $\chi_{N,\epsilon}$ almost identical for $N=2^{10}$ and $N=2^{11}$.) 

For model A, we observe a steady increase of  $\chi_{N,\epsilon}$ with $N$, compare  the horizontal dashed line indicating an error threshold $\epsilon=10^{-5}$ in Figs.~\ref{fig:tci_convergence}d) and f). To further access the efficiency of the algorithm, we can analyze the number $N_{\rm func}$ of function evaluations of the function $F(\bm \tau)$ in the construction of the MPS, which is needed to bring the error below a given threshold $\epsilon$, see Figs.~\ref{fig:tci_convergence}g) and h).  This should be compared to the  number of function evaluations needed to compute the integral directly on an equidistant time-ordered mesh, which is $N^3/6$ and $N^5/24$ for order $m=2$ and $m=3$, respectively (see dashed and solid black lines in Figs.~\ref{fig:tci_convergence}g) and h). The number $N_{\rm func}$ in the TCI integration is considerably smaller than in the direct integration (e.g., smaller by $8$-$10$ orders of magnitude at order $m=3$ and $N=2^{11}$). Moreover, in TCI, $N_{\rm func}$ increases less strongly with $N$ (a fit would give $N^{p}$ with $p$ around $1.5$). 

 We also remark that at fixed time the error of the integrals converges like $\mathcal{O}(h^{p})$ with the time-step $h$, where $p$ is determined by the quadrature rule used to compute the convolutions \eqref{retconv5}. We use a $5$th order Gregory quadrature rule \cite{Nessi}, and a convergence with $p=7$ is demonstrated in App.~\ref{FFT1}.

\begin{figure}[tbp]
\centerline{
\includegraphics[width=0.5\textwidth]{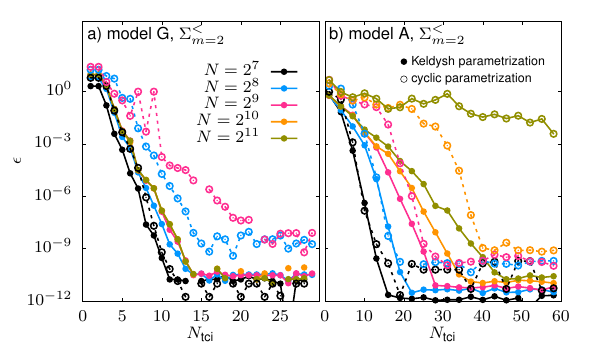}
}
\caption{
Analogous convergence analysis for the computation of the self-energy as in Fig.~\ref{fig:tci_convergence}, but for the lesser component $\Sigma^<$. Open and filled dots correspond to the cyclic parametrization and Keldsyh parametrization, respectively.
}
\label{tci_convergence3}
\end{figure}

The promising behavior of TCI for $\Sigma^>$  is similar to the finding for the imaginary time integrals \cite{Erpenbeck2023b}, which may be expected due to the similar structure of the algorithm. We can now proceed to the lesser component $\Sigma^<$, which needs a different representation of the integrals. In Fig.~\ref{tci_convergence3} we provide the analogous convergence analysis for $\Sigma^<$ for the same parameters as in Fig.~\ref{fig:tci_convergence}. In particular, we  compare the cyclic parametrization (open symbols) and the Keldysh parametrization (closed symbols). As expected from the discussion in Sec.~\ref{sec:celp}, the cyclic parametrization becomes worse for long times, when the decay of $\Delta(t)$ is fast compared to the maximum time $\tcut$. For the largest value of $\tcut$ in the simulation, where no data are shown for the cyclic parametrization, the error is simply out of the scale of the figure. Because in real simulations one will generally aim to choose $\tcut$ sufficiently large such that $\Delta$ is well localized within the interval $[-\tcut,\tcut]$, this unfavorable behavior of the cyclic parametrization could hinder a stable computation of the self-energy. For the Keldysh parametrization, instead, we observe that the bond dimension which is needed to reach a given accuracy does saturate with the length of the simulation interval (model G) or grow moderately (model A), similar for the greater component. For large $\tcut$, we therefore expect  the Keldysh parametrization to be generally favorable in spite of the additional cost due to the summation of the Keldysh indices $\bm \sigma$ in Eq.~\eqref{ksum}. For small times, however, both parametrizations perform similarly well, such that the numerically cheaper cyclic parametrization could be used if the integral can be restricted to a small interval, possibly in combination with suitable extrapolation techniques.

By comparison to Fig.~\ref{fig:tci_convergence} we note that even with the Keldysh parametrization, the required bond dimension is smaller for $\Sigma^>$ than for $\Sigma^<$ at the same order. This can be expected because for  $\Sigma^>$, TCI needs to factorize only the function $F_{\mathcal{D}}$ [Eq.~\eqref{ffunc}], while for the lesser component (in Keldysh parametrization) both $F$ and $W$ [Eq.~\eqref{wfunc}] are nontrivial. 

\subsection{Anderson model}
\label{sec:siam}

As a second test, we perform some convergence analysis of the TCI solver for the Anderson model, which provides a benchmark towards its application within non-equilibrium DMFT \cite{Aoki2014}. The Hamiltonian is given by $H = H_{\rm loc} + H_{\rm hyb}$, where the local part
\begin{align}
H_{\rm loc} 
&=U\Big(c_{\uparrow}^\dagger c_{\uparrow} -\frac{1}{2}\Big)\Big(c_{\downarrow}^\dagger c_{\downarrow}  -\frac{1}{2}\Big)
\end{align}
describes a Fermions in a single orbital with local interaction $U$, with creation operators $c_\sigma^\dagger$  and annihilation operators $c_\sigma$ for fermions with spin $\sigma$, and 
\begin{align}
H_{\rm hyb}
&=
\sum_{p,\sigma} 
\Big[
\epsilon_{p} a_{p,\sigma}^\dagger a_{p,\sigma}
+
\Big(
V_{p} a_{p,\sigma}^\dagger c_{\sigma}+h.c.
\Big)
\Big]
\end{align}
describes the hybridization of the orbital with a continuum of bath orbitals at energies $\epsilon_p$. Integrating out the bath degrees of freedom gives the action $\mathcal{S}=\mathcal{S}_{\rm loc}+\mathcal{S}_{\rm hyb}$, with 
\begin{align}
\mathcal{S}_{\rm hyb}
=
-
\sum_{\sigma}
\int dt dt' 
c_{\sigma}^\dagger(t) \Delta_\sigma(t,t') c_{\sigma}(t'),
\end{align}
depending on the hybridization function 
\begin{align}
\Delta_\sigma(t,t') =-i\sum_{p} |V_p|^2 \langle T_{\mathcal{C}}a_{p\sigma}(t)a_{p\sigma}^\dagger(t')\rangle.
\end{align}
Specifically, we adopt a smooth box density of states
\begin{align}
D(\epsilon)
=
\frac{\Gamma}{(1+\exp^{\nu(\epsilon-W)})(1+\exp^{-\nu(\epsilon-W)})},
\end{align}
with bandwidth $W$ and total hybridization strength $\Gamma$; $\nu$ is a parameter to smoothen the high energy edges of the spectrum. With this, the hybridization function at inverse temperature $\beta$ is given by
\begin{align}
\Delta_\sigma^<(t) &=i\int d\omega e^{-i\omega t}\,D(\epsilon) f_\beta (\omega),
\\
\Delta_\sigma^>(t) &=-i\int d\omega e^{-i\omega t}\,D(\epsilon) f_\beta (-\omega),
\end{align}
with the Fermi function $ f_\beta (\omega)=(1+e^{\beta\omega})^{-1}$.

\begin{figure}[tbp]
\centerline{\includegraphics[width=0.5\textwidth]{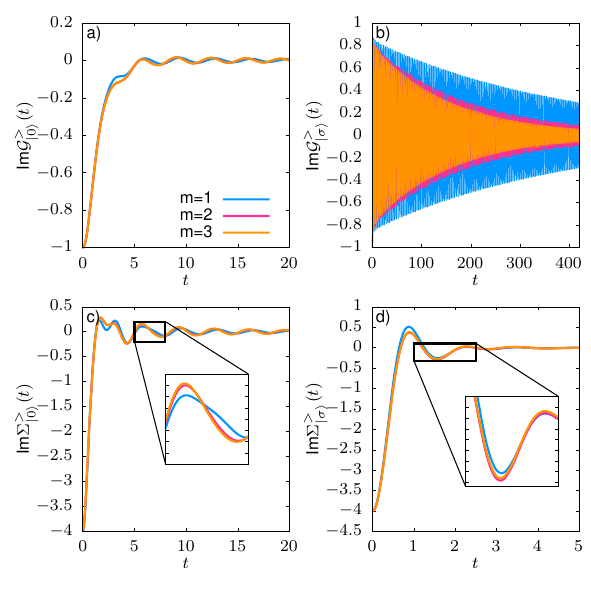}}
\caption{
Pseudo-particle Green's functions $\mathcal{G}^>(t)$ (upper panels) and self-energy $\Sigma^>(t)$  (lower panels) for the Anderson model, taking into account diagrams up to the indicated order $m$. Parameters are $U=6$, $W=10$, $\nu=3$, $\Gamma=0.2$. Left and right plots are for the zero particle sector $\{|0\rangle\}$ and  one particle sector $\{|\sigma\rangle\}$, respectively.}
\label{fig:siamppGs}
\end{figure}

In Fig.~\ref{fig:siamppGs} we exemplarily show the pseudo-particle Green's functions $\mathcal{G}^>(t)$ and self-energy $\Sigma^>(t)$ for a set of parameters in the Kondo regime. For the Anderson model, the  pseudo-particle Green's function $\mathcal{G}$ and self-energy $\Sigma$ is diagonal in the Fock basis of the impurity, and we denote by $\mathcal{G}_{|0\rangle}$ and $\mathcal{G}_{|\sigma\rangle}=\mathcal{G}_{|\uparrow\rangle}=\mathcal{G}_{|\downarrow\rangle}$ the components in the zero and one particle sector, respectively. Because of particle hole symmetry,  $\mathcal{G}_{|0\rangle}= \mathcal{G}_{|\uparrow\downarrow\rangle}$. For the parameters chosen here, the third order self-energy provides a small correction to the second order, while the lowest order $m_{\rm max}=1$ (NCA) is further off (compare also the physical spectral function below). 

An important feature to note, which is generic to the strong-coupling expansion, is that the propagators can decay much more slowly than the self-energy. This is in-particular seen in the one-particle sector, where the slow decay of $\mathcal{G}_{|\sigma\rangle}$ indicates the formation of long lived local moments (Figs.~\ref{fig:siamppGs}b) and d)). For this reason, the Dyson equation must be typically solved on a much longer  time window compared to the calculation of the integrals for the pseudo-particle Green's function $\Sigma$.  (For this purpose, the  $\mathcal{O}(N\log N)$ algorithm is useful for the solution of the Dyson equation is essential). The slow decay of $\mathcal{G}$ basically results from many self-energy insertions, i.e.,  high order disconnected diagrams for $\mathcal{G}$.

To access the convergence of the integrals in this typical setting, we now focus on the physical Green's function 
\begin{align}
G_{\rm imp}(t,t')=-i\langle T_{\mathcal{C}} c_\sigma (t)c_\sigma^\dagger (t')\rangle,
\end{align}
which is evaluated as explained in Secs.~\ref{sec:Cmain} and \ref{app:strcpC}. In order to have a controlled convergence test, we compute the $m$th order contributions $(G_{\rm imp}^>)_{m}$ and $(G_{\rm imp}^>)_{m}$ to the physical Green's functions with a given (non-self-consistent) input $\mathcal{G}$, for which we take the converged NCA result. Because the parameters are in a regime where the NCA result is at least qualitatively correct, this convergence analysis should be representative of the convergence of the integrals when $\mathcal{G}$ is given by the fully self-consistent solution.  

\begin{figure}[tbp]
\centerline{
\includegraphics[width=0.5\textwidth]{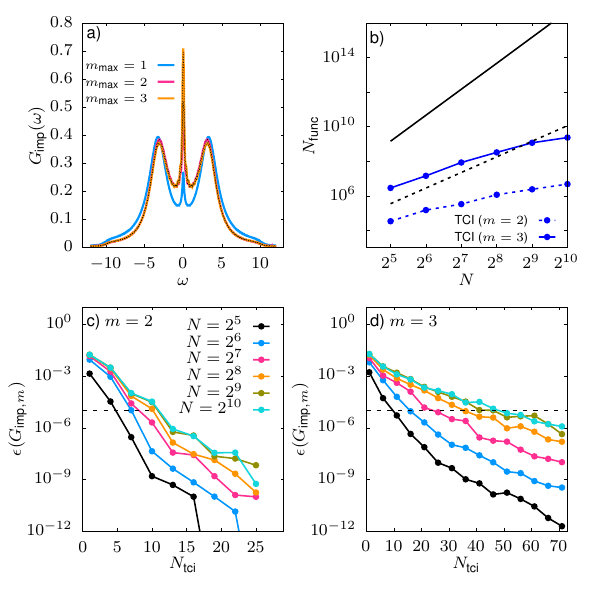}}
\caption{
a) Impurity spectral function $A_{\rm imp}(\omega)$ for a one-shot calculation, where the integrals for $G_{\rm imp}(t)$ are evaluated up to the indicated order $m_{\rm max}$, taking the converged NCA results as an input for $\mathcal{G}$. c) and d)  Convergence of the integrals for the 2nd order contribution $G_{\rm imp,m=2}$  (c) and the 3rd order contribution $G_{\rm imp,m=3}$   (d) with the number of TCI sweeps $N_{\rm tci}$. The different curves correspond to the evaluation of the  integral on time intervals of different length, where $N$ is the number of mesh points at fixed time-step $h=0.01$; see legend in c).  Only $G_{\rm imp}^>$ is analyzed, because the result for  $G_{\rm imp}^>$ is identical by particle hole symmetry. b) The number of function evaluations in the TCI, needed to suppress the error below a given threshold $\epsilon=10^{-5}$. Solid  and dashed black lines show the number of function evaluations for a direct integration at order $m=3$ and $m=2$, respectively.}
\label{fig:siam_Gimp_convergence}
\end{figure}

For illustration, Fig.~\ref{fig:siam_Gimp_convergence}a shows the spectral function $A_{\rm imp}(\omega)=G_{\rm imp}^>(\omega)-G_{\rm imp}^<(\omega)$ resulting from this one-shot calculation. One observes the typical three peak structure, with the Hubbard bands around $\omega=\pm U/2$ and a Kondo resonance at $\omega=0$. The Kondo resonance is enhanced when diagrams of higher order are included.  Figures~\ref{fig:siam_Gimp_convergence}c) and  Fig.~\ref{fig:siam_Gimp_convergence}d) then show the convergence analysis for the second order (c) and third order (d) contribution to the integrals for $G_{\rm imp}^>$, analogous to Fig.~\ref{fig:tci_convergence}, and Fig.~\ref{fig:siam_Gimp_convergence}b) illustrates the efficiency of the algorithm by evaluating the number $N_{\rm func}$ of function evaluations needed to evaluate the 2nd and 3rd order contributions to $G_{\rm imp}$ with a given accuracy. Similar to the toy model analysed in Fig.~\ref{fig:tci_convergence}, the TCI integration gets along with considerably fewer function evaluations. Also in this case, $N_{\rm func}$ increases with the length $N$ of the simulation interval less steeply than for the direct integration; compare the solid and dashed black lines in Fig.~\ref{fig:tci_convergence}d) with the TCI results. 

Because of the similar structure of the integrals, the convergence analysis for the lesser pseudo-particle self-energy $\Sigma^<$ is similar, while the integrals for $\Sigma^>$ converge faster, similar to the finding in the toy model (not shown here).

\subsection{Exactly solvable example: Holstein atom}
\label{sec:holsteinatom}

Finally, we study an exactly solvable model, which provides a nontrivial benchmark for the strong coupling expansion. Consider the simple Hamiltonian
\begin{align}
H = E c^\dagger c + \sum_{\alpha} \omega_\alpha b_\alpha^\dagger b_\alpha  + g\sum_{\alpha} (b_\alpha^\dagger + b_\alpha)c^\dagger c,
\label{hholstein}
\end{align}
which describes an isolated spinless Fermion orbital (with creation and annihilation operators $c$ and $c^\dagger$), coupled to a set of of bosonic modes with frequencies $\alpha$. The bosonic modes form a continuum with a given density of states $\Gamma(\omega) = \sum_{\alpha} \delta(\omega-\omega_\alpha)$ and the bosonic spectral function  $A(\omega) = \theta(\omega) \Gamma(\omega)-\theta(-\omega) \Gamma(-\omega)$.  The model can be used to  study of the effect of bosonic fluctuations on  Xray-absorbtion spectroscopy, where it basically replaces the simple cluster analysis of Ref.~\cite{Golez2024xas} with a continuum of modes. Here we use this model mainly as a an exactly solvable  benchmark, and defer a discussion of its physical properties, as well as a more detailed derivation of the exact solution, which uses a Lang-Firsov transformation similar to Ref.~\cite{Werner2007}, to a separate publication \cite{Paprotzki2024}.

By integrating out the bosonic degrees of freedoms, one obtains an action in the form $\mathcal{S}=\mathcal{S}_{\rm loc }+ \mathcal{S}_{\rm int}$, with the interaction in the form of Eq.~\eqref{intgenerega}, 
\begin{align} 
\mathcal{S}_{\rm int}&=-\frac{1}{2}\int dtdt'\, n(t)  \Delta(t,t') n(t'),
\end{align}
with a retarded density-density interaction. Here $n=c^\dagger c$, and  $\Delta(t,t')$ is related to the propagator of the bosonic bath,
\begin{align} 
\Delta^>(t)
=
-ig^2\int_{-\infty}^\infty d\omega A(\omega)
(1+b(\omega))e^{-i\omega t}
=
\Delta^<(-t).
 \end{align}
In this model, the state with zero electrons is  decoupled from the bath, so we can restrict the pseudo-particle Green's functions  and self-energies to the sector with $n=1$ fermion. The exact solution for the pseudo-particle Green's functions in this sector is given by 
 \begin{align}
\label{wfloewnwew}
\mathcal{G}^>(t) &=-ie^{-i\tilde E t } e^{F_b(t)}, \,\,\,\,\mathcal{G}^<(t)= i\rho_1 e^{-i\tilde E t } e^{F_b(-t)},
\end{align}
where $\rho_1$ is the occupation probability of the one fermion state, 
\begin{align}
F_b(t)=
 \int d\omega A(\omega) \frac{g^2}{\omega^2}
 (e^{-i\omega t}-1)(1+b(\omega)),
 \label{ceheaadeedlls}
\end{align}
and $\tilde E
= E-\int_0^\infty d\omega \Gamma(\omega)g^2/\omega$.
In the example here, we use a smooth density of states
\begin{align} 
\Gamma(\omega) = \frac{\omega^3}{\omega_c^4\cosh(\omega/\omega_c)},
\label{egwaaaadjds}
\end{align}
with a cutoff $\omega_c$, for which $\Delta E=\tilde E-E= \frac{g^2}{\omega_c}\frac{\pi^3}{8}$.

\begin{figure}[tbp]
\centerline{\includegraphics[width=0.5\textwidth]{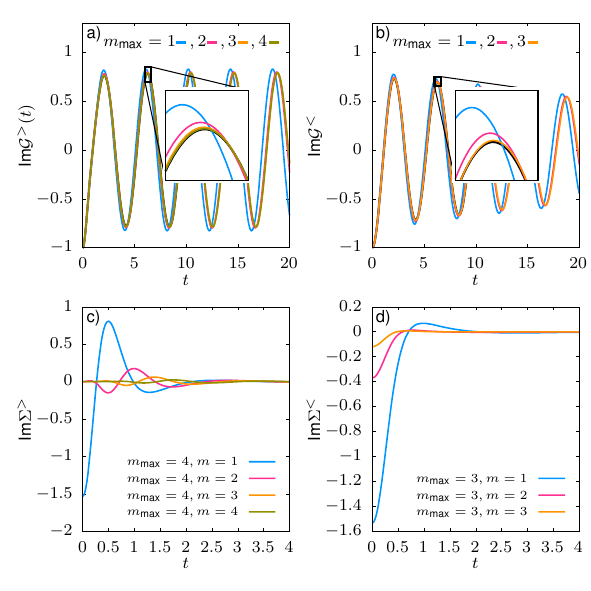}}
\caption{Propagator $\mathcal{G}^>(t)$ (a) and $\mathcal{G}^<(t)$ for the model \eqref{hholstein} with $E=2$, a density of states \eqref{egwaaaadjds} at $\omega_c=1$ and a coupling $g$ such that $\Delta E=0.5$. The inset shows a zoom in  a small region of the plot. Calculations are done including diagrams to up to the indicated order $m_{\rm max}$, using $N_{\rm tci}=40$; results for $N_{\rm tci}=30$ are indistinguishable of the scale of the plots. The decay of the lesser function with time is due to a regularization factor $\eta=0.02$ in the Dyson equation  (see App.~\ref{App:dyson}), which has been applied consistently to the exact solution in order to facilitate a comparison.
The lower plots c) and d) show the self energies $\Sigma_{m}$ at order $m$, for a simulation with maximum order $m_{\rm max}=4$ (c) and $m_{\rm max}=3$ (d).}
\label{fig:holsteon_atom}
\end{figure}

Figure \ref{fig:holsteon_atom} shows the pseudo-particle Green's functions for one given coupling strength. 
The exact $\mathcal{G}^>(t)$ shows a fast decay at early times, followed by a persistent oscillation at a frequency $\tilde E$ once the function $F_b$ has decayed to zero. Thus, the spectrum $\mathcal{G}(\omega)$ of the resolvent would show a $\delta$-peak at $\omega=\tilde E$, corresponding to an exact polaron state at the renormalized level energy $\tilde E$, and a continuum at higher energies. One can see that the strong coupling expansion converges order by order to this exact result, adding a certain confidence in the implementation.

\begin{figure}[tbp]
\centerline{\includegraphics[width=0.5\textwidth]{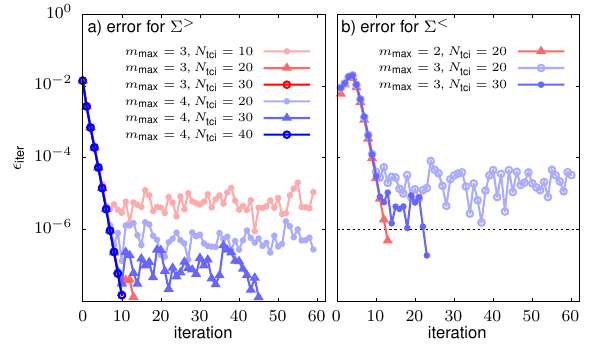}}
\caption{
Exemplary discussion of the convergence of the self-consistent iteration for $\Sigma^>$ (a) and $\Sigma^<$ (b), see text. The error $\epsilon_{\rm iter}$ is defined as the difference \eqref{norm2} between the self energy at two successive iterations. One can see a saturation of the convergence error at a level which is controlled by the accuracy of evaluation of the integrals, determined by $N_{\rm tci}$.}
\label{fig:holsteon_atim2}
\end{figure}

We can use this model also to demonstrate the behavior of the algorithm under a self-consistent solution of the Dyson equation with the numerically computed self energy.  In practice, the self-consistent solution is reached by an iteration, where the solution $\mathcal{G}^{(n)}$ of the Dyson equation \eqref{dyson} with a self-energy $\Sigma^{(n)}$  at iteration $n$ is used to determine a new self energy  $\Sigma^{(n)}_{\rm new}$;  the self-energy at the next iteration is then updated as $\Sigma^{(n+1)} = \alpha \Sigma^{(n)}_{\rm new}+(1-\alpha) \Sigma^{(n)}$, with a mixing factor $\alpha\in(0,1)$ to stabilize convergence ($\alpha\approx0.7$ for the present case). Fig.~\ref{fig:holsteon_atim2} shows the evolution of the convergence error $\epsilon_{\rm iter}$, where $\epsilon_{\rm iter}=||\Sigma^{(n)}-\Sigma^{(n)}_{\rm new}||_0$ is the difference between iterations, defined in analogy to Eq.~\eqref{norm2}. If the maximum bond dimension $N_{\rm tci}$ is too low, the iteration error settles at an level which controlled by the accuracy of the $\Sigma$ evaluation. Importantly, however, the iteration does not become unstable even at a relatively low $N_{\rm tci}$. With this, it can be in practice be a viable strategy to increase $N_{\rm tci}$ successively during the iteration process.

Finally, another interesting aspect of the model is the presence of the persistent oscillation in $\mathcal{G}^>$, which adds a subtlety to the solution within a steady state strong coupling approach: While the self-energy $\Sigma(t)$ decays rapidly with time, such that one can safely introduce a cutoff $\tcut$ for the integrals \eqref{SDegedws01}, the range of the convolution integral in the Dyson equation \eqref{dyson} is  only controlled by the decay of  $\mathcal{G}^>$.  The present model, with a non-decaying $\mathcal{G}$, requires a regularization of the Dyson equation, by introducing a suitable ``infinitesimal'' damping. Such a regularization should be added in a consistent way for $\mathcal{G}^<$ and $\mathcal{G}^>$, as described in App.~\ref{App:dyson}).

\section{Conclusion}
\label{sec:con}

In conclusion, we have discussed the evaluation of intermediate-order diagrams in the strong coupling expansion on the Keldysh contour, using TCI.  The key idea is that TCI can be used to factorize the nested part of the integrand in the diagrams, such that  the resulting integral to be computed as a recursion of convolution integrals. Possibly the most important technical aspect of the algorithm is the following: For a non-equilibrium steady state, one must obtain both the pseudo-particle self-energy  $\Sigma^>$, with time arguments on only branch of the Keldysh contour $\mathcal{C}$, and $\Sigma^<$, which has time arguments and integration variables on both branches of $\mathcal{C}$. For $\Sigma^>$, the time-ordered structure of the integrals is therefore similar to the evaluation of the strong coupling series on the imaginary-time contour, where TCI has been successfully used to compute integrals in the bare expansion of the pseudo-particle propagator \cite{Erpenbeck2023b}. For $\Sigma^<$ (as well as for the computation of physical correlation functions), we have discussed both a direct generalization of the time-ordered integrals for $\Sigma^>$ to contour-ordered integrals on a cyclically unfolded contour (``cyclic parametrization''), and a parametrization of the integrand in terms of physically ordered times, with an additional summation over Keldysh indices (``Keldysh parametrization''). We demonstrate and explain that the cyclic parametrization is less suitable to decompose an integrand if the  hybridization function has localized features in time, which will generally be the case if the time window for the integration is taken sufficiently large. This indicates that the Keldysh parametrization may in general be favorable to solve the strong-coupling expansion using TCI, and possibly also for other decomposition techniques (see below).  Alternatively, one could use the  numerically less costly cyclic parametrization, but try to limit the computation of integrals to a smaller time interval and use suitable extrapolation techniques.

We have tested the convergence of the integrals for several benchmark cases, namely a toy model with exponentially and algebraically decaying hybridization functions, as well as the Anderson-Holstein model and the Anderson model. For all cases, we find that TCI requires orders of magnitude fewer function evaluations to compute the integrals, as  compared to direct integration on an equidistant grid. This makes the present solver highly promising to improve non-equilibrium steady states DMFT simulations beyond the current state of the art NCA, but at a numerical cost below the numerically exact QMC \cite{Erpenbeck2023,Kunzel2024}. At the same time, it is important to keep in mind the  limitations of the proposed approach: 

(i) On the one hand the bond dimension of the matrix product state increases with dimension, such that simulations can become numerically costly even though they are much faster than the (inaccessible) direct evaluation of integrals. In this context, if would be interesting to also compare to alternative integration techniques such as quasi-Monte Carlo \cite{Strand2024,Bertrand2021,Macek2020} when generalized to steady-state non-equilibrium problems in the strong coupling formulation. (ii) More importantly, the number of skeleton diagrams grows factorially with the order. An possible route to reduce the number of diagrams at least at low order could be vertex re-summations of the strong coupling expansion \cite{Kim2022,Kim2023}. Alternatively, one may explore a possible use of TCI to sample also the space of diagrams.  In the form implemented in this manuscript, the evaluation of the strong-coupling expansion will reman restricted to low orders.  Nevertheless, for many applications of non-equilibrium DMFT, even the second or third order would represent a significant improvement over the existing NCA simulations, which can  in particularly become uncontrolled  for multi-orbital models. A more detailed analysis of self-consistent solutions in models with more than one orbital will be deferred to future work. 

Finally, one can draw an interesting link to an alternative approach to factorize the nested integrand, which is based on a decomposition of the hybridization function in terms of complex exponentials. The latter has been developed in Ref.~\cite{Kaye2024} for imaginary-time diagrams, and it leads to a similar recursive set of convolution integrals for each given set of exponentials in the decomposition.  In fact, if $\Delta$ and $\mathcal{G}$ can be expressed as a sum of at most $\chi$ exponentials, one can in principle construct an MPS such that the full integral takes the form  \eqref{nesterret}, and the trace over the MPS is identical as the summation over all sets of exponentials. In other words, the bond dimension of the MPS is limited by a power of $\chi$, depending on the structure of the function and number of $\Delta$-lines. It will be interesting to explore how a suitable generalization of the decomposition approach of Ref.~\cite{Kaye2024} to the real time  axis, and eventually to the two-branch Keldysh contour (for evaluating lesser components) compares to the factorization based on TCI.

\acknowledgements
I acknowledge fruitful conversations with Andre Erpenbeck, Denis Golez, Emanuel Gull,  Jason Kaye, Fabian Künzel, Eva Paprotzki, Olivier Parcollet, Hugo Strand, Xavier Waintal, and Philipp Werner. In particular, I thank Xavier Waintal for providing access to an early version of the tensor library \cite{Fernandez2024}, on which the TCI factorization in this manuscript is based. Funding is acknowledged through the Deutsche Forschungsgemeinschaft through  QUAST-FOR5249 -  449872909 (Project P6), and through the Cluster of Excellence „CUI: Advanced Imaging of Matter“ of the Deutsche Forschungsgemeinschaft (DFG) – EXC 2056 – project ID 390715994.

\begin{appendix}

\section{Steady state strong coupling expansion}
\label{app:strcpl}

In order to present a more self-contained manuscript, we briefly recapitulate the derivation of the strong coupling series within the Keldysh formalism. The presentation mainly follows the formulation in Refs.~\cite{Aoki2014} and \cite{Eckstein2010nca}, adapted to steady state problems. 

\subsection{Expansion of $Z$}

\begin{figure}[t]
\centerline{\includegraphics[width=0.4\textwidth]{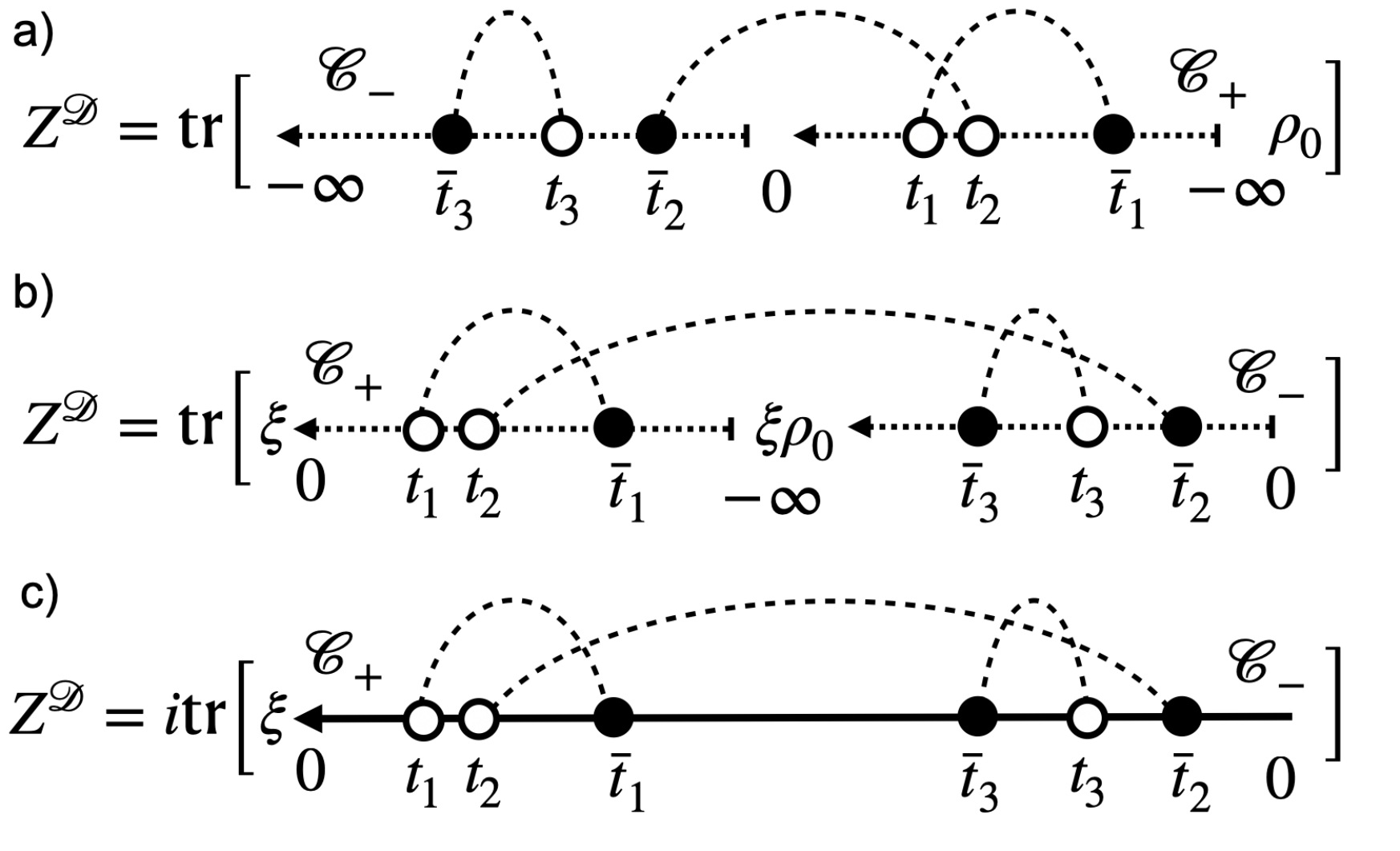}}
\caption{a) Diagrammatic representation of a term in the expansion \eqref{Zexpansionbare}. Full (empty) dots are vertices $\bar v$ ($v$). 
In c), we have made one $i$ factor explicit, such that we assotiate a sign $(-1)^{F'} i^{m}$ with the remainder of the diagram in Fig.~\ref{Zdiagram}c).}
\label{Zdiagram}
\end{figure}

All expectation values of impurity observables are evaluated by representing Eq.~\eqref{Cexpval} in a basis $\{|m\rangle\}$ of the local Hilbert space.  We first discuss the expansion of the normalization $Z=\text{tr} \big[\rho_0 T_{\mathcal{C}}  e^{i\mathcal{S}}\big]$. Expanding the term $e^{i\mathcal{S}_{\rm int}}$ under the contour order gives
\begin{align}
\text{tr}
&\Big[\rho_0
T_{\mathcal{C}}
 e^{i\mathcal{S}} \Big]
 =
 \sum_{m=0}^\infty
 \frac{(-i)^m}{m!}
  \sum_{\alpha_1,...,\alpha_m}  
   \int_\mathcal{C} d\bar t_1d t_1\cdots d\bar t_md t_m
\nonumber \\
 &\times\,\,\, \text{tr}\Big[\rho_0
 T_{\mathcal{C}}
 e^{i\mathcal{S}_{loc}}
 \bar v_{\alpha_1}(\bar t_1) v_{\alpha_1}( t_1)
 \cdots 
 \bar v_{\alpha_m} (\bar t_m) v_{\alpha_m}(t_m) 
 \Big]
 \nonumber\\
 &\times\,\,\, 
 \Delta_\alpha (\bar t_1,t_1)\cdots\Delta_\alpha (\bar t_m,t_m).
\label{Zexpansionbare}
\end{align}
After ordering the operators in the  trace term according to the contour order, the argument of the trace becomes a product of time-evolution operators and vertices. The full expression can therefore be represented as a sum over all diagrams of the form as shown in Fig.~\ref{Zdiagram}a), where the dotted (dashed) lines represent local time-evolution operators (interaction lines).  Each topology (defined by the directional interaction lines) is taken into account  with only one labelling of the vertices, due to the combinatorical factor $m!$ in Eq.~\eqref{Zexpansionbare}. The sign of the diagram is  $(-i)^m (-1)^f$, where the Fermion sign $f$ is given by the number of permutations of Fermi operators needed to bring the vertices $ \bar v_{\alpha_1} v_{\alpha_1} \cdots  \bar v_{\alpha_m}  \bar v_{\alpha_m}$ in the correct order;  $f$ can simply be evaluated as the number of crossing between Fermion interaction lines plus the number of Fermion lines which point opposite to the contour ordering (a fermionic interaction line is one that connects two Fermi operators). For example, if we assume that all interactions in Fig.~\ref{Zdiagram}a) are fermionic, there is one crossing (between $1$ and $2$) and one reversed lines ($1$). 

For later convenience, we then can permute the operators cyclically under the trace (Fig.~\ref{Zdiagram}b), such that the time ordering corresponds to the cyclic ordering introduced in the main text, which proceeds from $0_-$ to $-\infty$ on $\mathcal{C}_-$ and from $-\infty$ to $0_+$ on $\mathcal{C}_+$. We still want to evaluate the sign of the diagram by the simple rule  $(-i)^m (-1)^{f'}$ where now $f'$ is given by the number of fermion crossings and the number of lines which are against cyclic ordering. However, lines which connect $\mathcal{C}_+$ and $\mathcal{C}_-$ have opposite direction with respect to conventional order and cyclic order, such that  $(-i)^m (-1)^{f'}$ and  $(-i)^m (-1)^{f}$ differ by $(-1)^{f_{+-}}$, where $f_{+-}$ is the number of fermion lines connecting $\mathcal{C}_+$ and $\mathcal{C}_-$. To account for the difference, we simply insert the Fermion parity operator $\xi$, which is $-1$ $(+1)$ for states with an odd (even) number of fermions, at the two connection points between the two contour branches, as shown in Fig.~\ref{Zdiagram}b). Note that the  the statistics of pseudo-particles has no particular meaning, and instead of introducing the factor $\xi$ one could also just adapt the rules to determine the sign. The factor $\xi$ is kept to be consistent with earlier literature. 

To further make the analogy between time-evolution operators and contour Green's functions we define the noninteracting pseudo-particle Greensfunction $\mathcal{G}_{0}(t,t')$ on the two branch contour such that
\begin{align}
\label{appg0}
\mathcal{G}_{0}^>(t-t')&=-i\,\mathcal{U}_{\rm loc}(t,t'),
\\
\mathcal{G}_{0}^>(t-t')&=-i\,\mathcal{U}_{\rm loc}(t,t_0)\xi\rho_0\mathcal{U}_{\rm loc}(t_0,t'),
\label{g0000w}
\end{align}
where $\mathcal{U}_{\rm loc}(t,t')$ is the time evolution operator for the local Hamiltonian, $t_0\to-\infty$ is the initial time, and $\rho_0$ the initial density matrix; see Eq.~\eqref{lesgtrparam} for the definition of lesser and greater components of the  pseudo-particle Green's functions. Obviously, the hermitian property Eq.~\eqref{lesgtrparamherm} is satisfied. In the diagram, we can now replace the time-evolution operators by pseudo-particle Green's functions (solid lines in Fig.~\ref{Zdiagram}c). For a diagram of order $m$, there are $2m$ vertices and $2m+1$ $\mathcal{G}_0$ lines. The diagram then has an additional factor $i^{2m+1}$ to translate $\mathcal{G}_0$ back to time-evolution operators, giving an overall factor $(-1)^{f'} i^{m+1}$. 

To sum the series of diagrams one can use standard partial resummation techniques: The interacting pseudoparticle self-energy  $\Sigma(t,t')$ of order $m$ for $t\succ t'$ is defined by all diagrams as shown in Fig.~\ref{Sdiagram}, which are irreducible in the $\mathcal{G}_0$-lines, and a sign $(-1)^{f'} i^{m}$. With this, the solution of the integral  Dyson equation Eq.~\eqref{dyson} generates  the dressed propagator $\mathcal{G}(t,t')$, which up to a factor $-i$ is the full evolution operator with all possible interaction lines inserted between $t\succ t'$.  The full partition function is given determined by the evolution operator along the full contour,
\begin{align}
Z
=
i\text{tr}[
\xi \mathcal{G}(0_+,0_-)],
\end{align}
where the factor $i$ is again due to the relation \eqref{appg0} between evolution operators and propagators. Furthermore, one can add one level of self-consistency and keep only skeleton diagrams in the expansion (i.e., diagrams in which none of the lines has a self-energy insertion), while replacing the $\mathcal{G}_0$ lines in the diagrams with the fully dressed lines. This leads to the self-consistent formalism described in Sec.~\ref{sec:mainstrc} of the main text.

\subsection{Self-energy diagrams}

For the numerical evaluation of the strong-coupling expansion, we need to generate all self-energy diagrams at a given order.  A self-energy diagrams of order $m$ (c.f.~Fig.~\ref{Sdiagram}) is labeled by a list 
\begin{align}
\label{Dlist}
\mathcal{D}=(\bar a_m,a_m;\alpha_m),...,(\bar a_1,a_1,\alpha_1), 
\end{align}
where $\bar a_j$ ($a_j$) are the labels of the time arguments for the endpoint (start-point) of the line $j$, and $\alpha_j$ is the interaction flavor, i.e.,  line $j$ directs from a vertex $v_{\alpha_j}$ at time $t_{a_{j}}$ to a vertex $\bar  v_{\alpha_j}$ at time $t_{\bar a_{j}}$ in a diagram where times $t_{2m}\succ t_{2m-1} \succ\cdots\succ t_{1}$ are cyclically ordered. Since our algorithm aims at low orders, we can explicitly construct a list of all possible diagrams. Even though the number of diagrams grows rapidly with order, it is the evaluation of the diagrams which sets the numerical effort, and not the determination of this list. We therefore construct the list of diagrams by a simple exclusion algorithm:

\begin{itemize}
\item[(1)]
Determine all groupings of the numbers $(n,...,1)$ into  pairs $(r_m,s_m)\cdots(r_2,s_2)(r_1,s_1)$, where  $s_m> \cdots>s_2>s_1=1$ and $r_j>s_j$. For example, the diagram Fig.~\ref{Sdiagram} corresponds to $(8,5)(7,3)(6,2)(4,1)$. 

\item[(2)] 
For each diagram, check whether it has a sub-cluster of interconnected adjacent vertices. If it does, the diagram is either not skeleton-like or not one-particle irreducible, and will be disregarded. We find $1$ skeleton topology at order $m=1$, $1$ at $m=2$, $4$ at $m=3$, $27$ at $m=4$, $248$ at $m=5$, $2830$ at $m=6$, $38232$ at $m=7$, and $593859$ at $m=8$.

\item[(3)]
For each non-directed skeleton topology, generate all $2^m$ directed topologies, by allowing for each line both directions $(r_j,s_j)$ and $(s_j,r_j)$. For example, the diagram of Fig.~\ref{Sdiagram} corresponds to the list $(5,8)(7,6)(2,6)(4,1)$, where now the right (left) entry in the pair corresponds to the closed (open) circle, i.e., the operator $\bar v$ ($v$). 

\item[(4)]
For each line, choose all possible types of interaction. This results in the list \eqref{Dlist}. We denote the operators at the time-points $t_n,..,t_1$ by $v^{\mathcal{D}}_{n},...,v^{\mathcal{D}}_{1}$, and keep only those diagrams for which the product $v^{\mathcal{D}}_{n} \cdots v^{\mathcal{D}}_{1}$ is non-vanishing (e.g., this excludes diagrams with two successive fermion creation of annihilation operators).

\item[(5)]
Finally, we determine the Fermion sign for the diagram: If we retain only the fermion interaction lines $(a,b)(c,d),...,(e,f)$. The sign is $(-1)^P$, where $P$ is the number of permuting the indices to bring these numbers in increasing order. The overall factor for the diagram $\mathcal{D}$ is $C_{\mathcal{D}}(-1)^P i^m$.
\end{itemize}
The self-energy diagram then has the analytical expression \eqref{SDegedws01} stated in the main text. A simpler case are symmetric interaction lines, corresponding to a term 
\begin{align}
\mathcal{S}_{\text{int}}
=
-\frac{1}{2}\int_{\mathcal{C}} d\bar t dt\,
v(\bar t) \Delta (\bar t,t) v(t)
\label{intgeneregasym}
\end{align}
in the action, with a symmetric function $ \Delta (\bar t,t)= \Delta (t,\bar t) $ connecting identical (bosonic) operators $v$. An example is the boson-induced density-density interaction in the electron boson  model of Sec.~\ref{sec:holsteinatom}. In this case, forward and backward lines give the same contribution with a weight $\Delta/2$, such that we can omit step (3) above and take into account only forward lines with a weight $\Delta$.

\begin{figure}[tbp]
\centerline{\includegraphics[width=0.4\textwidth]{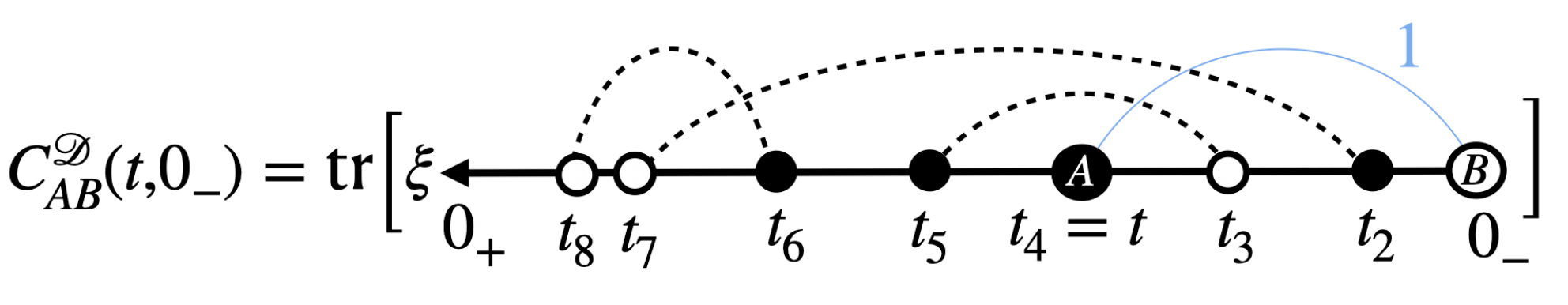}}
\caption{A diagram for the correlation function $C_{AB}(t,t')$ ($t \succ t'$) of order $m=4$. The light blue line is the virtual line $\Delta=1$ connecting the external vertices.}
\label{Gdiagram}
\end{figure}

\subsection{Expectation values and correlation functions}
\label{app:strcpC}

The expectation value of an operator $A$ is obtained by inserting $A$ at time $0$ in the contour-ordered expectation value \eqref{Cexpval}, and expanding the action. After transforming to the cyclic order, this gives
\begin{align}
\langle A \rangle=  
\frac{i}{Z}
\text{tr}[
A \xi \mathcal{G}(0_+,0_-)]
=
\frac{i}{Z}
\text{tr}[
A \xi \mathcal{G}^<(0)]
\label{pseudoexp}
\end{align}
with the normalization
\begin{align}
Z=  i
\text{tr}[
\xi \mathcal{G}(0_+,0_-)]=
i\text{tr}[
\xi \mathcal{G}^<(0)].
\label{norm}
\end{align}
Here $\mathcal{G}(0_+,0_-)$ is the dressed propagator along the full contour, which contains all interaction insertions. Expression \eqref{pseudoexp} also implies that $i\xi \mathcal{G}^<(0)/Z$ is the reduced density matrix of the impurity site. The normalization $Z$ can be considered as a free quantity, as all lesser quantities in the self-consistent steady state solution can be rescaled with an arbitrary factor (see also the discussion in Sec.~\eqref{appdysgen}). Usually, we choose the normalization $Z=1$.

The diagrams for two-point  correlation functions
\begin{align}
C_{AB}(t,t')=-i\langle T_{\mathcal{C} }A(t) B(t') \rangle
\label{Cab}
\end{align}
have additional vertices $A$ and $B$ on the contour. Because of time-translational invariance, we can choose without $t'=0_-$, and $t\succ t'$. One can repeat all steps in the derivation for the diagrams for $Z$, adding the overall  factor $(-i)$and the two vertices $A$ and $B$. The correlation function $C_{AB}$ is then given by all diagrams as shown in Fig.~\ref{Gdiagram}), which can be re-summed into skeleton diagrams by removing the self-energy insertions from the internal $\mathcal{G}$ lines. The sign is given by  a factor $(-i)^m$ from the action, an overall factor $-i$ from the definition of $C_{AB}$, a factor $i^{2m+2}$ from translating Green's functions in evolution operators, and a factor $(-1)^f$, where $f$ is the number of fermion interaction lines with anti-cyclic order and the number of fermion line crossings. If $A$ and $B$ are fermionic, we count the connection from $B$ to $A$ as an additional fermion line. The diagram thus has a factor $-i^{m+1}(-1)^{f}$.

In the following we will call a $C$-diagram to be of order $m$ if it has $m-1$ interaction lines. In this way, all $C$-diagrams of order $m$ are obtained by constructing all $\Sigma$ diagrams at order $m$ for which the line connecting to the first vertex in cyclic order is a forward pointing line connecting a vertex $B$ at its starting point  and a vertex $A$ at its end.  The same result can also be viewed as the result of taking the derivative of $Z$ with respect to the function $c(t,t')$ after adding a source term $\mathcal{S}_{c} = -\int_{\mathcal{C}} d\bar t dt\, A(\bar t) c (\bar t,t) B(t) $ to the action. 

The analytical expression for a diagram $\mathcal{D}$ with vertices $t_n\succ\cdots t_2\succ t_1=0_-$, in which $t_s$ is the external argument (i.e., the end-point of the line connecting $B$ and $A$) is 
\begin{align}
&C_{\mathcal{D}}
(t_s,0_-) 
= 
\int_{0_+\succ t_n \succ \cdots t_2 \succ 0_-} 
\hspace*{-19mm}
dt_n\cdots dt_{s+1} dt_{s-1}\cdots dt_2
\nonumber\\
&\times \,\,\, \text{tr}[\xi \mathcal{G}(0_+,t_n) v^{\mathcal{D}}_{n} \mathcal{G}(t_n,t_{n-1})v^{\mathcal{D}}_{n-1} \cdots v^{\mathcal{D}}_2 \mathcal{G}(t_2,t_{1}) v^{\mathcal{D}}_1]
\nonumber\\
&\times \,\,\,
C_{\mathcal{D}} \prod_{j=1}^m \Delta_{\alpha_j}(t_{\bar a_j},t_{a_j}),
  \label{CDegedws01}
\end{align}
where the external vertices are $v^{\mathcal{D}}_1=B$ and $v^{\mathcal{D}}_s=A$, and the factor $\Delta$ for the line connecting $A$ and $B$ is simply $1$. The  contribution of order $m=1$ to the correlation function (NCA contribution), which has no integration, is given by 
\begin{align}
&C_{AB}(t,0_-)= i \text{tr}\big[\xi \mathcal{G}(0_+,t)A\mathcal{G}(t,0_-)B\big].
\label{cnca}
\end{align}

The integrals  \eqref{CDegedws01} are solved, analogous to self-energy diagrams, in Keldysh parametrization
\begin{align}
C_{\mathcal{D}}^{\kappa}(r_s)
=
\sum_{\bm \sigma}{}' \int_{\bm r | r_{s(\bm\sigma)}}  \!\! d \bm r \,\,
I^C_{\bm \sigma,\mathcal{D}}(\bm r) \delta_{\kappa,\sigma_s} S_{\bm \sigma,\mathcal{D}}
\end{align}
where $I^C_{\bm \sigma,\mathcal{D}}(\bm r)$ is the integrand of \eqref{CDegedws01} written in terms of physically ordered times $\bm r$,
$\sum_{\bm \sigma}'$  sums over all set of Keldysh indices with $\sigma_1=-1$, $S_{\bm \sigma,\mathcal{D}}=(-1)^l$ with $l$ being the number of integration vertices on $\mathcal{C}_-$, $s$ is the external time argument (depending on $\bm \sigma$ and $\mathcal{D}$), and $  \delta_{\kappa,\sigma_s} $ selects the Keldysh index of the external vertex $s$ to be $-1 (+1)$ when $\kappa$ is $>(<)$, respectively. The integral 
is solved as for the lesser self-energy, by a factorization of the integrand in terms of physical time differences.

\section{Evaluation of the time-ordered convolutions Eq.~\eqref{retconv5} using FFT}
\label{FFT1}

Here we state the detailed expressions for the evaluation of the convolution integral
\begin{align}
c(\tau)= \int_{0}^{\tau} d\tau' a(\tau-\tau')b(\tau'),
\label{gdehjwjzssddd}
\end{align}
using FFT in combination with higher order accurate quadrature rules. We assume that $\tau$ runs on a grid of $N=2^M$ points
\begin{align}
 \tau \in m h, 
 \,\,\,
 m\in\{0,1,...,N-1\}, \,\,\,\,(N-1)h=\tcut,
\end{align}
and we denote $x_m=x(mh)$ for $x=a,b,c$.  To evaluate the integral, we use a $k$th order quadrature rule; $k=0$ would correspond to the Trapez rule. The general quadrature for the integral \eqref{gdehjwjzssddd} used below has the form 
\begin{align}
\label{tconv01A}
c_n = 
\begin{cases}
h \sum_{m=0}^{n} w^{(n)}_{m} a_{n-m} b_m  & n\ge k
\\
h \sum_{m,l=0}^{k} w^{(n)}_{m,l}  a_{m} b_l  & n< k
\end{cases},
\end{align}
with quadrature weights $w^{(n)}_{m}$. The second case ($n< k$) takes care of the fact that the $k$th order accurate integral needs $k+1$ points: In this case we use a boundary convolution as described in Chapter 8.6 of Ref.~\cite{Nessi}, which computes the integral from a polynomial interpolation of $a$ and $b$ on the interval $[0,kh]$.  In the generic case $(n\ge k)$, we use a  Gregory quadrature rule as described in Chapter 8.5 of Ref.~\cite{Nessi}.  Most importantly, for the Gregory quadrature all weights are unity apart from the boundary weights $w^{(n)}_{m}$ with $m\le k$ or $n-m\le k$. We can thus write
\begin{align}
 \label{tconv01A2}
c_n = h \sum_{m=0}^{n}  a_{n-m} b_m + h \delta c_n,
 \end{align}
 where $\delta c$ is a  boundary correction that can be evaluated in effort $\mathcal{O}(N)$; $\delta c_n$ is implicitly defined by the difference of Eq.~\eqref{tconv01A} and the first term on the right hand side of Eq.~\eqref{tconv01A2}. For a Trapez rule, with $k=0$, one would simply have  $ \delta c_n=\tfrac12(a_0b_n+a_nb_0)$.  
 
To evaluate the remaining sum in \eqref{tconv01A2}, one can use a standard trick to bring the term in a form which can be solved using FFT: We extend the vectors to a time interval of double length $\bar N=2N$, with 
 \begin{align}
\bar x_{n}
=
\begin{cases}
x_{n} & n=0,...,N-1
\\
0 & n=N,...,2N-1
\end{cases},
\label{retardedoperiod}
 \end{align}
for  $x=a,b$. If the vectors with an overbar are be understood to be extended to arbitrary $n$ with a periodicity of $\bar N$, $\bar x_{n\pm\bar N} = \bar  x_n$, the sum in \eqref{tconv01A2} corresponds to the unrestricted  convolution,
\begin{align}
\bar c_n \equiv  \sum_{m=0}^{n}  a_{n-m} b_m = \sum_{m=0}^{\bar N-1}  \bar a_{n-m} \bar b_m   \text{~~~for~~}0\le n\le N-1.
\nonumber
 \end{align}
The unrestricted convolution is evaluated using FFT
 \begin{align}
&\tilde x_{k}
=
\sum_{n=0}^{\bar N-1}  \bar x_{n} e^{i 2\pi kn  /\bar N} \,\,\,\,k=0,...,\bar N-1
\label{forwardfft}
 \end{align}
for $x=a,b$, and
 \begin{align}
\bar c_{n}
=
\frac{1}{\bar N}
\sum_{k=0}^{\bar N-1} \tilde a_{k}
\tilde b_{k}
e^{-i 2\pi kn /\bar N},
 \end{align}
so that $c_n=\delta c_n + h \bar c_n$  for $n=0,...,\bar N-1$.  The values $ \bar c_n$ for $n=N,...,2N-1$ can be disregarded. The two Fourier sums are evaluated in $\mathcal{O}(\bar N\log \bar N)$ steps. 

\begin{figure}[tbp]
\centerline{
\includegraphics[width=0.5\textwidth]{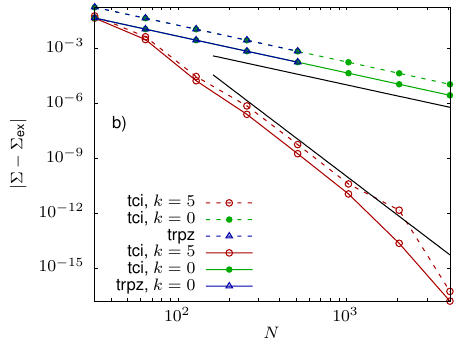}
}
\caption{
Convergence of the integrals for $\Sigma$ for model G (Sec.~\ref{sec:convan}) at order $m=2$ with decreasing time-step (see text for explanation). In addition, we show results for a simple trapezoidal integration of the integral \eqref{SDegedws01}, labelled ``trpz''. The latter is done only for smaller $N$ as simulations become much slower than the TCI evaluation.  Solid black lines indicate the behavior $\epsilon\sim h^3$ (trapez) and $\epsilon \sim h^7$ (tci).
}
\label{fig:h_convergence}
\end{figure}

To demonstrate the convergence of the integrals with the timestep $h$, we focus on ``model G'' of Sec.~\ref{sec:convan}. The result of the convergence analysis is shown in Fig.~\eqref{fig:h_convergence}.  We compute $\Sigma^>$ and $\Sigma^<$ at order $m=2$ using a fine mesh of $N=N_{\text{max}}=2^{12}$ points with step $h=h_{\rm min}=0.2/2^7$. We then decrease $N$ in powers of $2$ down to $N=2^{5}$ (increase $h$ in powers to $2$ up to  $h=2^{7}h_{\rm min}$). The error for $N<N_{\text{max}}$ is evaluated by the difference \eqref{norm2} to the most accurate result on the  grid of $2^5$ points which is common to all simulations. For integration order of order $k=0$, the TCI integration coincides with the trapzezoidal integration, for which the error decreases as $h^3$. For the $5$th order Gregory quadrature, we observe a convergence with a higher power $h^7$.

\section{Dyson equation}
\label{App:dyson}

\subsection{General considerations} 
\label{appdysgen}
In this section, we describe the numerical solution of the pseudo-particle Dyson equation \eqref{dyson}. Because the bare time-evolution operators \eqref{g0000w} satisfy the Schr\"odinger equation with $H_{\rm loc}$, the integral Dyson equation is equivalent to the integral differential equation
\begin{align}
(i\partial_t - H_{\rm loc})\mathcal{G}(t,t') -\int_{t\succ t_1 \succ t'}\!\!\!\! dt_1  \Sigma( t,t_1)\mathcal{G}(t_1,t')
=0,
\end{align}
which must be solved with the initial condition $\mathcal{G}(t,t')=-i$  for $t\to t'$. The Dyson equation for the greater component is obtained, e.g., by choosing both $t $ and $t'$ on $\mathcal{C}_+$, with $t>0$  and $t'=0$:
\begin{align}
\label{dysonret}
(i\partial_t - H_{\rm loc})\mathcal{G}^>(t) - \int_0^t d\bar t \,\Sigma^>(t-\bar t)\mathcal{G}^>(\bar t)=0,
\end{align}
with the initial condition $\mathcal{G}^>(0)=-i$. Conversely, the differential equation for $\mathcal{G}^<(t)$ at $t<0$ is obtained, e.g., by choosing $t<0$  on $\mathcal{C}_+$ and  $t'=0$ on $\mathcal{C}_-$,
\begin{align}
\label{dysonles}
(i\partial_t - H_{\rm loc})\mathcal{G}^<(t) - &\int_{-\infty}^t  dt_1 \Sigma^>(t-\bar t)\mathcal{G}^<(\bar t)
\nonumber\\
&=
 -\int_{-\infty}^0 d\bar t \,\Sigma^<(t-\bar t)\mathcal{G}^>(\bar t).
\end{align}
Using Eq.~\eqref{dysonret} for $\mathcal{G}^>$ (or rather its conjugate), Eq.~\eqref{dysonles} is transformed into an integral equation
\begin{align}
\mathcal{G}^<(t) 
&=
-\int_{-\infty}^t dt_1
\int_{-\infty}^0 dt_2
\mathcal{G}^>(t-t_1)
\Sigma^<(t_1-t_2)
\mathcal{G}^>(t_2)
\nonumber
\\
&+
\lim_{\bar t\to-\infty}
\mathcal{G}^>(t-\bar t)
\mathcal{G}^<(\bar t).
\label{whwjwla071}
\end{align}
Generically, we assume that the  interacting propagator $\mathcal{G}^>(t)$ decays for large times, such that the boundary term in Eq.~\eqref{whwjwla071} can be neglected. In this case, Eqs.~\eqref{whwjwla071} and \eqref{dysonret} can be solved on a sufficiently long time window, as described in Ref.~\cite{Li2021}, or in real time at a similar numerical cost $\mathcal{O}(N\log N)$ using $k$th-order accurate integration routines (see Sec.~\ref{dysonrealtimme}).

In some cases, however, $\mathcal{G}^>$ does not decay at all, or decays only very slowly. A simple example is the Holstein atom of Sec.~\ref{sec:holsteinatom}, where $\mathcal{G}^>$ has persistent oscillations at long times, or the single impurity model with a gapped spectral function. In this case, one could separate out the asymptotic  behavior of $\mathcal{G}^<$, either by an exponential fit of the tail or looking for the zeros in the inverse $\mathcal{G}^>(\omega+i0)^{-1} = [\omega+i0- H_{\rm loc} - \Sigma^>(\omega+i0)]$ in frequency space. Then the singular part of the integral \eqref{whwjwla071} could be combined exactly with the boundary term, to derive an equation for the singular (non-decaying) contribution to $\mathcal{G}^<$. 

The more conventional approach is to introduce an additional damping factor in $\mathcal{G}^>$. In this case, however, some case should be taken to introduce a consistent contribution to the lesser self-energy. A small damping can be physically introduced in a physical way by inserting a small self-energy due to an artificial interaction 
\begin{align} 
\label{seta}
\mathcal{S}_{\eta}&=-\frac{\eta}{2}\int dtdt'\,\hat 1(t)  \Delta(t,t') \hat 1(t'),
 \end{align}
 where $\eta$ is a positive number, $\hat 1$ the identity matrix, and $\Delta^>(t)=\Delta^<(t) = -i 2\delta_{\rm sym}(t)$
 a symmetric function which is arbitrary short range in time (``symmetric delta function''). For such an instantaneous interaction, only the first order self-energy is nonzero:
\begin{align} 
\Sigma_{\eta}^>(t-t') &= -i2\eta \delta_{\rm sym}(t-t'),
\\
\Sigma_{\eta}^< (t-t') &= 2\eta \delta_{\rm sym}(t-t') \mathcal{G}^<(0).
 \end{align}
Note that  $\mathcal{G}^<(0)\sim -i\xi\hat \rho$ is proportional to the local density matrix by Eq~\eqref{pseudoexp}]. Replacing $\Sigma$ in Eqs.~\eqref{dysonret} and \eqref{whwjwla071} with $\Sigma+\Sigma_\eta$ gives rise to the following equations:
 \begin{align}
 \label{dysonwithetaret}
&(i\partial_t - H_{loc} + i\eta )\mathcal{G}(t,t') -\int_{t\succ t_1 \succ t'}\!\!\!\! dt_1  \Sigma( t,t_1)\mathcal{G}(t_1,t')
=0,
\\
&\mathcal{G}^<(t) 
=
-\int_{-\infty}^t dt_1
\int_{-\infty}^0 dt_2
\mathcal{G}^>(t-t_1)
\Sigma^<(t_1-t_2)
\mathcal{G}^>(t_2)
\nonumber
\\
&
\,\,\,\,\,\,\,\,\,\,\,\,\,\,\,\,\,\,\,\,-
\eta 
\int_{-\infty}^t dt_1
\mathcal{G}^>(t-t_1)
2 \mathcal{G}^<(0)
\mathcal{G}^>(t_1)
\label{whwjwla073}.
\end{align}
The equations can be solved using the same real time techniques as without the $\eta$ regularization. 
In equilibrium, where propagators and self-energies should satisfy a fluctuation dissipation relation $\mathcal{G}^<(\omega) = e^{-\beta\omega} \mathcal{G}^>(\omega)$, in can be more convenient to introduce the regulator as a frequency dependent quantity which establishe the fluctuation dissipation relation for $\Sigma_{\eta}$  \cite{Li2021}.

We note that in both Eq.~\ref{whwjwla071} and Eq.~\eqref{whwjwla073} , the right-hand side depends linearly on $\mathcal{G}^<$, since $\Sigma$ itself is linearly dependent on $\mathcal{G}^<$. As a consequence, the solution can be rescaled arbitrarily without affecting the physical results. (The factor \eqref{norm} appears only as a normalization constant in expectation values.) Typically, we normalize $\mathcal{G}^<$ such that $Z = 1$. Rescaling can be incorporated during the self-consistent iteration process, or applied after convergence. However, in the latter case, the norm may become exponentially large or small during the iterations. 


\subsection{Dyson equation for $\mathcal{G}^>$ in real time}
\label{dysonrealtimme}

While the Dyson equation can be solved in frequency space \cite{Li2021} with an effort $\mathcal{O}(N \log N)$, the solution  a real-time, as implemented in NESSi \cite{Nessi}, is more accurate (integrals are computed using high order accurate integrators) but comes with a higher effort $\mathcal{O}(N^2)$. In the following we explain how a $k$th order accurate solution implemented in NESSi \cite{Nessi}  is achieved in $\mathcal{O}(N \log N)$ effort using FFT, by exploiting the translational invariance of the self-energy. A combination of an $\mathcal{O}(N \log N)$ algorithm with boundary weights for high order accurate integration has also been used in Ref.~\cite{Kaye2023b} for the computation of the left-mixing Green's function in the equilibrium Dyson equation.

We start with Eq.~\eqref{dysonwithetaret} for the greater component in a discretized form, using the notation $G_j\equiv \mathcal{G}^>(jh)$.  We first assume that we have determined $G_0$,..., $G_{p-1}$ for some initial $p$ steps ($p\ge k$) using the regular retarded Dyson equation from NESSi. To solve the equation for the remaining steps, NESSi uses a $(k+1)$-order backward differentiation formula  $(\partial_t G)_j \approx \frac{1}{h}\sum_{l=0}^{k+1} \alpha_l G_{j-l}$ in combination with a $k$-order Gregory quadrature, which for $n>k$ takes the form 
\begin{align}
&
[\Sigma^> \ast \mathcal{G}^>]_{n} = \sum_{m=0}^k w_m^{(n)} \Sigma_{n-m} G_m
\nonumber
\\
&\,\,\,\,+
\sum_{m=k+1}^{n-k-1} \Sigma_{n-m} G_m
+
\sum_{m=n-k}^{n} w_{n-m} \Sigma_{n-m} G_m, 
\end{align}
with quadrature weights $w_{n-m}$ at the upper boundary that do not depend on $n$; for the weight factors $\alpha_l$ and $w_{m}^{(n)}$, see Ref.~\cite{Nessi}. The discretized Eq.~\eqref{dysonwithetaret} then becomes a linear equation for an unknown vector $X_m=G_{p+m}$, $m=0,...,N-1$, with the triangular form
\begin{align}
\label{linear}
\sum_{m=0}^n M_{n-m} X_{m} = \phi_{n},
\end{align}
where 
\begin{align}
M_{l}
=
\begin{cases}
i\alpha_0/h-H_{\rm loc}+i\eta  - h w_0 \Sigma_0 & \text{~for~} l=0
\\
i\alpha_l/h - h w_{l}  \Sigma_l & \text{~for~} l=1,...,k
\\
i\alpha_l/h -   h\Sigma_l & \text{~for~} l=k+1
\\
-   h\Sigma_l & \text{~for~} l>k+1
\end{cases},
\nonumber
\end{align}
while the right hand side of \eqref{linear}  depends on the initial values $(G_0,G_{1},...,G_{p-1})$,
\begin{align}
\phi_{n}
\!=\!
\sum_{m=0}^{p-1}
\big[
hw^{(p+n)}_m \Sigma_{p+n-m}G_{m} 
-
\frac{i}{h}
\alpha_{p+n-m}
\big]G_{m}, 
\end{align}
where in the last term it is understood that $\alpha_l=0$ for $l>k+1$.
To solve the linear equation, we proceed analogous to the convolution integral \eqref{gdehjwjzssddd}, and extend the size of the problem from $N$ to $\bar N=2N$, with the extension
 \begin{align}
\bar x_{n}
=
\begin{cases}
x_{n} & n=0,...,N-1
\\
0 & n=N,...,2N-1
\end{cases},
 \end{align}
for  $x=X,\phi$, and the periodic extension $\bar M_{n\pm\bar N} = \bar  M_n$ for $M$. The solution of the extended system
\begin{align}
\label{linear-extended}
\sum_{m=0}^{\bar N} \bar M_{n-m} \bar X_{m} = \bar \phi_{n},
\end{align}
is then identical to the solution of \eqref{linear} of the steps $n=0,...N-1$, provided that the solution $\bar X_{n}=0$ for $n=N,...,\bar N-1$. To achieve the latter condition, one must choose $N$ sufficiently large  such that the solution of the differential equation is exponentially small at the upper boundary (either because of the intrinsic decay due to the self energy $\Sigma$, or due to the $\eta$ factor). Because Eq.~\eqref{linear-extended} has the form of an unrestricted convolution, it can be solved in $\mathcal{O}(\bar N \log \bar N)$ effort using FFT. 

\subsection{Dyson equation for $\mathcal{G}^<$ in real time}
\label{dysonrealtimme}

For the lesser Dyson equation \eqref{whwjwla073}, we assume that due to the decay of the (possibly regularized) $\mathcal{G}$ the range of the integrals can be restricted to $[-\tcut,0]$, with $\tcut=(N-1)h$. The double integral in Eq.~\eqref{whwjwla073} is then reduced to two successive convolutions,
\begin{align}
C_1(t) &=
-\int_{-\tcut}^0 dt_2
\Sigma^<(-t-t_2)
\mathcal{G}^>(t_2)
\label{convGG}
\end{align}
for $t\in [0,\tcut]$, and 
\begin{align}
\mathcal{G}^<(-t) &=
\int_{-\tcut}^{-t} dt_1
\mathcal{G}^>(-t-t_1)
C_1(-t_1)
\\
&\stackrel{\tau=-t-t_1}{=}
\int_0^{\tcut-t} d\tau\,
\mathcal{G}^>(\tau)
C_1(\tau+t).
\label{geheB04}
\end{align}
These are convolutions, which both can be treated similarly to the retarded convolution in Sec.~\ref{FFT1} by using an FFT convolution on an extended interval, adding boundary corrections for higher order quadrature rules at $\mathcal{O}(N)$ cost.

\end{appendix}


\begin{thebibliography}{53}%
\makeatletter
\providecommand \@ifxundefined [1]{%
 \@ifx{#1\undefined}
}%
\providecommand \@ifnum [1]{%
 \ifnum #1\expandafter \@firstoftwo
 \else \expandafter \@secondoftwo
 \fi
}%
\providecommand \@ifx [1]{%
 \ifx #1\expandafter \@firstoftwo
 \else \expandafter \@secondoftwo
 \fi
}%
\providecommand \natexlab [1]{#1}%
\providecommand \enquote  [1]{``#1''}%
\providecommand \bibnamefont  [1]{#1}%
\providecommand \bibfnamefont [1]{#1}%
\providecommand \citenamefont [1]{#1}%
\providecommand \href@noop [0]{\@secondoftwo}%
\providecommand \href [0]{\begingroup \@sanitize@url \@href}%
\providecommand \@href[1]{\@@startlink{#1}\@@href}%
\providecommand \@@href[1]{\endgroup#1\@@endlink}%
\providecommand \@sanitize@url [0]{\catcode `\\12\catcode `\$12\catcode
  `\&12\catcode `\#12\catcode `\^12\catcode `\_12\catcode `\%12\relax}%
\providecommand \@@startlink[1]{}%
\providecommand \@@endlink[0]{}%
\providecommand \url  [0]{\begingroup\@sanitize@url \@url }%
\providecommand \@url [1]{\endgroup\@href {#1}{\urlprefix }}%
\providecommand \urlprefix  [0]{URL }%
\providecommand \Eprint [0]{\href }%
\providecommand \doibase [0]{https://doi.org/}%
\providecommand \selectlanguage [0]{\@gobble}%
\providecommand \bibinfo  [0]{\@secondoftwo}%
\providecommand \bibfield  [0]{\@secondoftwo}%
\providecommand \translation [1]{[#1]}%
\providecommand \BibitemOpen [0]{}%
\providecommand \bibitemStop [0]{}%
\providecommand \bibitemNoStop [0]{.\EOS\space}%
\providecommand \EOS [0]{\spacefactor3000\relax}%
\providecommand \BibitemShut  [1]{\csname bibitem#1\endcsname}%
\let\auto@bib@innerbib\@empty
\bibitem [{\citenamefont {Georges}\ \emph {et~al.}(1996)\citenamefont
  {Georges}, \citenamefont {Kotliar}, \citenamefont {Krauth},\ and\
  \citenamefont {Rozenberg}}]{Georges1996}%
  \BibitemOpen
  \bibfield  {author} {\bibinfo {author} {\bibfnamefont {A.}~\bibnamefont
  {Georges}}, \bibinfo {author} {\bibfnamefont {G.}~\bibnamefont {Kotliar}},
  \bibinfo {author} {\bibfnamefont {W.}~\bibnamefont {Krauth}},\ and\ \bibinfo
  {author} {\bibfnamefont {M.~J.}\ \bibnamefont {Rozenberg}},\ }\bibfield
  {title} {\bibinfo {title} {Dynamical mean-field theory of strongly correlated
  fermion systems and the limit of infinite dimensions},\ }\href
  {https://doi.org/10.1103/RevModPhys.68.13} {\bibfield  {journal} {\bibinfo
  {journal} {Rev. Mod. Phys.}\ }\textbf {\bibinfo {volume} {68}},\ \bibinfo
  {pages} {13} (\bibinfo {year} {1996})}\BibitemShut {NoStop}%
\bibitem [{\citenamefont {Gull}\ \emph {et~al.}(2011)\citenamefont {Gull},
  \citenamefont {Millis}, \citenamefont {Lichtenstein}, \citenamefont
  {Rubtsov}, \citenamefont {Troyer},\ and\ \citenamefont {Werner}}]{Gull2011}%
  \BibitemOpen
  \bibfield  {author} {\bibinfo {author} {\bibfnamefont {E.}~\bibnamefont
  {Gull}}, \bibinfo {author} {\bibfnamefont {A.~J.}\ \bibnamefont {Millis}},
  \bibinfo {author} {\bibfnamefont {A.~I.}\ \bibnamefont {Lichtenstein}},
  \bibinfo {author} {\bibfnamefont {A.~N.}\ \bibnamefont {Rubtsov}}, \bibinfo
  {author} {\bibfnamefont {M.}~\bibnamefont {Troyer}},\ and\ \bibinfo {author}
  {\bibfnamefont {P.}~\bibnamefont {Werner}},\ }\bibfield  {title} {\bibinfo
  {title} {Continuous-time {Monte} {Carlo} methods for quantum impurity
  models},\ }\href {https://doi.org/10.1103/RevModPhys.83.349} {\bibfield
  {journal} {\bibinfo  {journal} {Rev. Mod. Phys.}\ }\textbf {\bibinfo {volume}
  {83}},\ \bibinfo {pages} {349} (\bibinfo {year} {2011})}\BibitemShut
  {NoStop}%
\bibitem [{\citenamefont {Werner}\ \emph {et~al.}(2009)\citenamefont {Werner},
  \citenamefont {Oka},\ and\ \citenamefont {Millis}}]{Werner2009}%
  \BibitemOpen
  \bibfield  {author} {\bibinfo {author} {\bibfnamefont {P.}~\bibnamefont
  {Werner}}, \bibinfo {author} {\bibfnamefont {T.}~\bibnamefont {Oka}},\ and\
  \bibinfo {author} {\bibfnamefont {A.~J.}\ \bibnamefont {Millis}},\ }\bibfield
   {title} {\bibinfo {title} {Diagrammatic {Monte} {Carlo} simulation of
  nonequilibrium systems},\ }\href@noop {} {\bibfield  {journal} {\bibinfo
  {journal} {Phys. Rev. B}\ }\textbf {\bibinfo {volume} {79}},\ \bibinfo
  {pages} {035320} (\bibinfo {year} {2009})}\BibitemShut {NoStop}%
\bibitem [{\citenamefont {Cohen}\ \emph {et~al.}(2014)\citenamefont {Cohen},
  \citenamefont {Gull}, \citenamefont {Reichman},\ and\ \citenamefont
  {Millis}}]{Cohen2014}%
  \BibitemOpen
  \bibfield  {author} {\bibinfo {author} {\bibfnamefont {G.}~\bibnamefont
  {Cohen}}, \bibinfo {author} {\bibfnamefont {E.}~\bibnamefont {Gull}},
  \bibinfo {author} {\bibfnamefont {D.~R.}\ \bibnamefont {Reichman}},\ and\
  \bibinfo {author} {\bibfnamefont {A.~J.}\ \bibnamefont {Millis}},\ }\bibfield
   {title} {\bibinfo {title} {Green's functions from real-time bold-line
  {Monte} {Carlo} calculations: Spectral properties of the nonequilibrium
  {Anderson} impurity model},\ }\href
  {https://doi.org/10.1103/PhysRevLett.112.146802} {\bibfield  {journal}
  {\bibinfo  {journal} {Phys. Rev. Lett.}\ }\textbf {\bibinfo {volume} {112}},\
  \bibinfo {pages} {146802} (\bibinfo {year} {2014})}\BibitemShut {NoStop}%
\bibitem [{\citenamefont {Cohen}\ \emph {et~al.}(2015)\citenamefont {Cohen},
  \citenamefont {Gull}, \citenamefont {Reichman},\ and\ \citenamefont
  {Millis}}]{Cohen2015}%
  \BibitemOpen
  \bibfield  {author} {\bibinfo {author} {\bibfnamefont {G.}~\bibnamefont
  {Cohen}}, \bibinfo {author} {\bibfnamefont {E.}~\bibnamefont {Gull}},
  \bibinfo {author} {\bibfnamefont {D.~R.}\ \bibnamefont {Reichman}},\ and\
  \bibinfo {author} {\bibfnamefont {A.~J.}\ \bibnamefont {Millis}},\ }\bibfield
   {title} {\bibinfo {title} {Taming the dynamical sign problem in real-time
  evolution of quantum many-body problems},\ }\href
  {https://doi.org/10.1103/PhysRevLett.115.266802} {\bibfield  {journal}
  {\bibinfo  {journal} {Phys. Rev. Lett.}\ }\textbf {\bibinfo {volume} {115}},\
  \bibinfo {pages} {266802} (\bibinfo {year} {2015})}\BibitemShut {NoStop}%
\bibitem [{\citenamefont {Wolf}\ \emph {et~al.}(2014)\citenamefont {Wolf},
  \citenamefont {McCulloch},\ and\ \citenamefont {Schollw\"ock}}]{Wolf2014}%
  \BibitemOpen
  \bibfield  {author} {\bibinfo {author} {\bibfnamefont {F.~A.}\ \bibnamefont
  {Wolf}}, \bibinfo {author} {\bibfnamefont {I.~P.}\ \bibnamefont
  {McCulloch}},\ and\ \bibinfo {author} {\bibfnamefont {U.}~\bibnamefont
  {Schollw\"ock}},\ }\bibfield  {title} {\bibinfo {title} {Solving
  nonequilibrium dynamical mean-field theory using matrix product states},\
  }\href {https://doi.org/10.1103/PhysRevB.90.235131} {\bibfield  {journal}
  {\bibinfo  {journal} {Phys. Rev. B}\ }\textbf {\bibinfo {volume} {90}},\
  \bibinfo {pages} {235131} (\bibinfo {year} {2014})}\BibitemShut {NoStop}%
\bibitem [{\citenamefont {Gramsch}\ \emph {et~al.}(2013)\citenamefont
  {Gramsch}, \citenamefont {Balzer}, \citenamefont {Eckstein},\ and\
  \citenamefont {Kollar}}]{Gramsch2013}%
  \BibitemOpen
  \bibfield  {author} {\bibinfo {author} {\bibfnamefont {C.}~\bibnamefont
  {Gramsch}}, \bibinfo {author} {\bibfnamefont {K.}~\bibnamefont {Balzer}},
  \bibinfo {author} {\bibfnamefont {M.}~\bibnamefont {Eckstein}},\ and\
  \bibinfo {author} {\bibfnamefont {M.}~\bibnamefont {Kollar}},\ }\bibfield
  {title} {\bibinfo {title} {{Hamiltonian}-based impurity solver for
  nonequilibrium dynamical mean-field theory},\ }\href
  {https://doi.org/10.1103/PhysRevB.88.235106} {\bibfield  {journal} {\bibinfo
  {journal} {Phys. Rev. B}\ }\textbf {\bibinfo {volume} {88}},\ \bibinfo
  {pages} {235106} (\bibinfo {year} {2013})}\BibitemShut {NoStop}%
\bibitem [{\citenamefont {Cohen}\ and\ \citenamefont
  {Rabani}(2011)}]{Cohen2011}%
  \BibitemOpen
  \bibfield  {author} {\bibinfo {author} {\bibfnamefont {G.}~\bibnamefont
  {Cohen}}\ and\ \bibinfo {author} {\bibfnamefont {E.}~\bibnamefont {Rabani}},\
  }\bibfield  {title} {\bibinfo {title} {Memory effects in nonequilibrium
  quantum impurity models},\ }\href
  {https://doi.org/10.1103/PhysRevB.84.075150} {\bibfield  {journal} {\bibinfo
  {journal} {Phys. Rev. B}\ }\textbf {\bibinfo {volume} {84}},\ \bibinfo
  {pages} {075150} (\bibinfo {year} {2011})}\BibitemShut {NoStop}%
\bibitem [{\citenamefont {Park}\ \emph {et~al.}(2024)\citenamefont {Park},
  \citenamefont {Ng}, \citenamefont {Reichman},\ and\ \citenamefont
  {Chan}}]{Park2024}%
  \BibitemOpen
  \bibfield  {author} {\bibinfo {author} {\bibfnamefont {G.}~\bibnamefont
  {Park}}, \bibinfo {author} {\bibfnamefont {N.}~\bibnamefont {Ng}}, \bibinfo
  {author} {\bibfnamefont {D.~R.}\ \bibnamefont {Reichman}},\ and\ \bibinfo
  {author} {\bibfnamefont {G.~K.-L.}\ \bibnamefont {Chan}},\ }\bibfield
  {title} {\bibinfo {title} {Tensor network influence functionals in the
  continuous-time limit: Connections to quantum embedding, bath discretization,
  and higher-order time propagation},\ }\href
  {https://doi.org/10.1103/PhysRevB.110.045104} {\bibfield  {journal} {\bibinfo
   {journal} {Phys. Rev. B}\ }\textbf {\bibinfo {volume} {110}},\ \bibinfo
  {pages} {045104} (\bibinfo {year} {2024})}\BibitemShut {NoStop}%
\bibitem [{\citenamefont {Thoenniss}\ \emph
  {et~al.}(2023{\natexlab{a}})\citenamefont {Thoenniss}, \citenamefont
  {Lerose},\ and\ \citenamefont {Abanin}}]{Thoenniss2023b}%
  \BibitemOpen
  \bibfield  {author} {\bibinfo {author} {\bibfnamefont {J.}~\bibnamefont
  {Thoenniss}}, \bibinfo {author} {\bibfnamefont {A.}~\bibnamefont {Lerose}},\
  and\ \bibinfo {author} {\bibfnamefont {D.~A.}\ \bibnamefont {Abanin}},\
  }\bibfield  {title} {\bibinfo {title} {Nonequilibrium quantum impurity
  problems via matrix-product states in the temporal domain},\ }\href
  {https://doi.org/10.1103/PhysRevB.107.195101} {\bibfield  {journal} {\bibinfo
   {journal} {Phys. Rev. B}\ }\textbf {\bibinfo {volume} {107}},\ \bibinfo
  {pages} {195101} (\bibinfo {year} {2023}{\natexlab{a}})}\BibitemShut
  {NoStop}%
\bibitem [{\citenamefont {Thoenniss}\ \emph
  {et~al.}(2023{\natexlab{b}})\citenamefont {Thoenniss}, \citenamefont
  {Sonner}, \citenamefont {Lerose},\ and\ \citenamefont
  {Abanin}}]{Thoenniss2023}%
  \BibitemOpen
  \bibfield  {author} {\bibinfo {author} {\bibfnamefont {J.}~\bibnamefont
  {Thoenniss}}, \bibinfo {author} {\bibfnamefont {M.}~\bibnamefont {Sonner}},
  \bibinfo {author} {\bibfnamefont {A.}~\bibnamefont {Lerose}},\ and\ \bibinfo
  {author} {\bibfnamefont {D.~A.}\ \bibnamefont {Abanin}},\ }\bibfield  {title}
  {\bibinfo {title} {Efficient method for quantum impurity problems out of
  equilibrium},\ }\href {https://doi.org/10.1103/PhysRevB.107.L201115}
  {\bibfield  {journal} {\bibinfo  {journal} {Phys. Rev. B}\ }\textbf {\bibinfo
  {volume} {107}},\ \bibinfo {pages} {L201115} (\bibinfo {year}
  {2023}{\natexlab{b}})}\BibitemShut {NoStop}%
\bibitem [{\citenamefont {Aoki}\ \emph {et~al.}(2014)\citenamefont {Aoki},
  \citenamefont {Tsuji}, \citenamefont {Eckstein}, \citenamefont {Kollar},
  \citenamefont {Oka},\ and\ \citenamefont {Werner}}]{Aoki2014}%
  \BibitemOpen
  \bibfield  {author} {\bibinfo {author} {\bibfnamefont {H.}~\bibnamefont
  {Aoki}}, \bibinfo {author} {\bibfnamefont {N.}~\bibnamefont {Tsuji}},
  \bibinfo {author} {\bibfnamefont {M.}~\bibnamefont {Eckstein}}, \bibinfo
  {author} {\bibfnamefont {M.}~\bibnamefont {Kollar}}, \bibinfo {author}
  {\bibfnamefont {T.}~\bibnamefont {Oka}},\ and\ \bibinfo {author}
  {\bibfnamefont {P.}~\bibnamefont {Werner}},\ }\bibfield  {title} {\bibinfo
  {title} {Nonequilibrium dynamical mean-field theory and its applications},\
  }\href {https://doi.org/10.1103/RevModPhys.86.779} {\bibfield  {journal}
  {\bibinfo  {journal} {Rev. Mod. Phys.}\ }\textbf {\bibinfo {volume} {86}},\
  \bibinfo {pages} {779} (\bibinfo {year} {2014})}\BibitemShut {NoStop}%
\bibitem [{\citenamefont {Lange}\ \emph {et~al.}(2017)\citenamefont {Lange},
  \citenamefont {Lenar{\v{c}}i{\v{c}}},\ and\ \citenamefont
  {Rosch}}]{Lange2017}%
  \BibitemOpen
  \bibfield  {author} {\bibinfo {author} {\bibfnamefont {F.}~\bibnamefont
  {Lange}}, \bibinfo {author} {\bibfnamefont {Z.}~\bibnamefont
  {Lenar{\v{c}}i{\v{c}}}},\ and\ \bibinfo {author} {\bibfnamefont
  {A.}~\bibnamefont {Rosch}},\ }\bibfield  {title} {\bibinfo {title} {Pumping
  approximately integrable systems},\ }\href
  {https://doi.org/10.1038/ncomms15767} {\bibfield  {journal} {\bibinfo
  {journal} {Nat. Commun.}\ }\textbf {\bibinfo {volume} {8}},\ \bibinfo {pages}
  {15767} (\bibinfo {year} {2017})}\BibitemShut {NoStop}%
\bibitem [{\citenamefont {Murakami}\ \emph {et~al.}(2023)\citenamefont
  {Murakami}, \citenamefont {Golež}, \citenamefont {Eckstein},\ and\
  \citenamefont {Werner}}]{Murakami2023}%
  \BibitemOpen
  \bibfield  {author} {\bibinfo {author} {\bibfnamefont {Y.}~\bibnamefont
  {Murakami}}, \bibinfo {author} {\bibfnamefont {D.}~\bibnamefont {Golež}},
  \bibinfo {author} {\bibfnamefont {M.}~\bibnamefont {Eckstein}},\ and\
  \bibinfo {author} {\bibfnamefont {P.}~\bibnamefont {Werner}},\ }\href@noop {}
  {\bibinfo {title} {Photo-induced nonequilibrium states in {Mott} insulators}}
  (\bibinfo {year} {2023}),\ \Eprint {https://arxiv.org/abs/2310.05201}
  {arXiv:2310.05201 [cond-mat.str-el]} \BibitemShut {NoStop}%
\bibitem [{\citenamefont {de~la Torre}\ \emph {et~al.}(2021)\citenamefont
  {de~la Torre}, \citenamefont {Kennes}, \citenamefont {Claassen},
  \citenamefont {Gerber}, \citenamefont {McIver},\ and\ \citenamefont
  {Sentef}}]{Sentef2021}%
  \BibitemOpen
  \bibfield  {author} {\bibinfo {author} {\bibfnamefont {A.}~\bibnamefont
  {de~la Torre}}, \bibinfo {author} {\bibfnamefont {D.~M.}\ \bibnamefont
  {Kennes}}, \bibinfo {author} {\bibfnamefont {M.}~\bibnamefont {Claassen}},
  \bibinfo {author} {\bibfnamefont {S.}~\bibnamefont {Gerber}}, \bibinfo
  {author} {\bibfnamefont {J.~W.}\ \bibnamefont {McIver}},\ and\ \bibinfo
  {author} {\bibfnamefont {M.~A.}\ \bibnamefont {Sentef}},\ }\bibfield  {title}
  {\bibinfo {title} {Colloquium: Nonthermal pathways to ultrafast control in
  quantum materials},\ }\href {https://doi.org/10.1103/RevModPhys.93.041002}
  {\bibfield  {journal} {\bibinfo  {journal} {Rev. Mod. Phys.}\ }\textbf
  {\bibinfo {volume} {93}},\ \bibinfo {pages} {041002} (\bibinfo {year}
  {2021})}\BibitemShut {NoStop}%
\bibitem [{\citenamefont {Li}\ and\ \citenamefont {Eckstein}(2021)}]{Li2021}%
  \BibitemOpen
  \bibfield  {author} {\bibinfo {author} {\bibfnamefont {J.}~\bibnamefont
  {Li}}\ and\ \bibinfo {author} {\bibfnamefont {M.}~\bibnamefont {Eckstein}},\
  }\bibfield  {title} {\bibinfo {title} {Nonequilibrium steady-state theory of
  photodoped {Mott} insulators},\ }\href
  {https://doi.org/10.1103/PhysRevB.103.045133} {\bibfield  {journal} {\bibinfo
   {journal} {Phys. Rev. B}\ }\textbf {\bibinfo {volume} {103}},\ \bibinfo
  {pages} {045133} (\bibinfo {year} {2021})}\BibitemShut {NoStop}%
\bibitem [{\citenamefont {Murakami}\ \emph {et~al.}(2022)\citenamefont
  {Murakami}, \citenamefont {Takayoshi}, \citenamefont {Kaneko}, \citenamefont
  {Sun}, \citenamefont {Golez}, \citenamefont {Millis},\ and\ \citenamefont
  {Werner}}]{Murakami2022}%
  \BibitemOpen
  \bibfield  {author} {\bibinfo {author} {\bibfnamefont {Y.}~\bibnamefont
  {Murakami}}, \bibinfo {author} {\bibfnamefont {S.}~\bibnamefont {Takayoshi}},
  \bibinfo {author} {\bibfnamefont {T.}~\bibnamefont {Kaneko}}, \bibinfo
  {author} {\bibfnamefont {Z.}~\bibnamefont {Sun}}, \bibinfo {author}
  {\bibfnamefont {D.}~\bibnamefont {Golez}}, \bibinfo {author} {\bibfnamefont
  {A.~J.}\ \bibnamefont {Millis}},\ and\ \bibinfo {author} {\bibfnamefont
  {P.}~\bibnamefont {Werner}},\ }\bibfield  {title} {\bibinfo {title}
  {Exploring nonequilibrium phases of photo-doped {Mott} insulators with
  generalized {Gibbs} ensembles},\ }\href
  {https://doi.org/10.1038/s42005-021-00799-7} {\bibfield  {journal} {\bibinfo
  {journal} {Commun. Phys.}\ }\textbf {\bibinfo {volume} {5}},\ \bibinfo
  {pages} {23} (\bibinfo {year} {2022})}\BibitemShut {NoStop}%
\bibitem [{\citenamefont {Li}\ \emph {et~al.}(2020)\citenamefont {Li},
  \citenamefont {Golez}, \citenamefont {Werner},\ and\ \citenamefont
  {Eckstein}}]{Li2020eta}%
  \BibitemOpen
  \bibfield  {author} {\bibinfo {author} {\bibfnamefont {J.}~\bibnamefont
  {Li}}, \bibinfo {author} {\bibfnamefont {D.}~\bibnamefont {Golez}}, \bibinfo
  {author} {\bibfnamefont {P.}~\bibnamefont {Werner}},\ and\ \bibinfo {author}
  {\bibfnamefont {M.}~\bibnamefont {Eckstein}},\ }\bibfield  {title} {\bibinfo
  {title} {$\ensuremath{\eta}$-paired superconducting hidden phase in
  photodoped {Mott} insulators},\ }\href
  {https://doi.org/10.1103/PhysRevB.102.165136} {\bibfield  {journal} {\bibinfo
   {journal} {Phys. Rev. B}\ }\textbf {\bibinfo {volume} {102}},\ \bibinfo
  {pages} {165136} (\bibinfo {year} {2020})}\BibitemShut {NoStop}%
\bibitem [{\citenamefont {Ray}\ \emph {et~al.}(2023)\citenamefont {Ray},
  \citenamefont {Murakami},\ and\ \citenamefont {Werner}}]{Ray2023}%
  \BibitemOpen
  \bibfield  {author} {\bibinfo {author} {\bibfnamefont {S.}~\bibnamefont
  {Ray}}, \bibinfo {author} {\bibfnamefont {Y.}~\bibnamefont {Murakami}},\ and\
  \bibinfo {author} {\bibfnamefont {P.}~\bibnamefont {Werner}},\ }\href@noop {}
  {\bibinfo {title} {Non-thermal superconductivity in photo-doped multi-orbital
  {Hubbard} systems}} (\bibinfo {year} {2023}),\ \Eprint
  {https://arxiv.org/abs/2305.06760} {arXiv:2305.06760 [cond-mat.str-el]}
  \BibitemShut {NoStop}%
\bibitem [{\citenamefont {Li}\ \emph {et~al.}(2023)\citenamefont {Li},
  \citenamefont {M\"uller}, \citenamefont {Kim}, \citenamefont {L\"auchli},\
  and\ \citenamefont {Werner}}]{Li2023}%
  \BibitemOpen
  \bibfield  {author} {\bibinfo {author} {\bibfnamefont {J.}~\bibnamefont
  {Li}}, \bibinfo {author} {\bibfnamefont {M.}~\bibnamefont {M\"uller}},
  \bibinfo {author} {\bibfnamefont {A.~J.}\ \bibnamefont {Kim}}, \bibinfo
  {author} {\bibfnamefont {A.~M.}\ \bibnamefont {L\"auchli}},\ and\ \bibinfo
  {author} {\bibfnamefont {P.}~\bibnamefont {Werner}},\ }\bibfield  {title}
  {\bibinfo {title} {Twisted chiral superconductivity in photodoped frustrated
  {Mott} insulators},\ }\href {https://doi.org/10.1103/PhysRevB.107.205115}
  {\bibfield  {journal} {\bibinfo  {journal} {Phys. Rev. B}\ }\textbf {\bibinfo
  {volume} {107}},\ \bibinfo {pages} {205115} (\bibinfo {year}
  {2023})}\BibitemShut {NoStop}%
\bibitem [{\citenamefont {Arrigoni}\ \emph {et~al.}(2013)\citenamefont
  {Arrigoni}, \citenamefont {Knap},\ and\ \citenamefont {von~der
  Linden}}]{Arrigoni2013}%
  \BibitemOpen
  \bibfield  {author} {\bibinfo {author} {\bibfnamefont {E.}~\bibnamefont
  {Arrigoni}}, \bibinfo {author} {\bibfnamefont {M.}~\bibnamefont {Knap}},\
  and\ \bibinfo {author} {\bibfnamefont {W.}~\bibnamefont {von~der Linden}},\
  }\bibfield  {title} {\bibinfo {title} {Nonequilibrium dynamical mean-field
  theory: An auxiliary quantum master equation approach},\ }\href
  {https://doi.org/10.1103/PhysRevLett.110.086403} {\bibfield  {journal}
  {\bibinfo  {journal} {Phys. Rev. Lett.}\ }\textbf {\bibinfo {volume} {110}},\
  \bibinfo {pages} {086403} (\bibinfo {year} {2013})}\BibitemShut {NoStop}%
\bibitem [{\citenamefont {Erpenbeck}\ \emph {et~al.}(2024)\citenamefont
  {Erpenbeck}, \citenamefont {Blommel}, \citenamefont {Zhang}, \citenamefont
  {Lin}, \citenamefont {Cohen},\ and\ \citenamefont {Gull}}]{Erpenbeck2024}%
  \BibitemOpen
  \bibfield  {author} {\bibinfo {author} {\bibfnamefont {A.}~\bibnamefont
  {Erpenbeck}}, \bibinfo {author} {\bibfnamefont {T.}~\bibnamefont {Blommel}},
  \bibinfo {author} {\bibfnamefont {L.}~\bibnamefont {Zhang}}, \bibinfo
  {author} {\bibfnamefont {W.-T.}\ \bibnamefont {Lin}}, \bibinfo {author}
  {\bibfnamefont {G.}~\bibnamefont {Cohen}},\ and\ \bibinfo {author}
  {\bibfnamefont {E.}~\bibnamefont {Gull}},\ }\bibfield  {title} {\bibinfo
  {title} {{Steady-state properties of multi-orbital systems using quantum
  {Monte} {Carlo}}},\ }\href {https://doi.org/10.1063/5.0226253} {\bibfield
  {journal} {\bibinfo  {journal} {The Journal of Chemical Physics}\ }\textbf
  {\bibinfo {volume} {161}},\ \bibinfo {pages} {094104} (\bibinfo {year}
  {2024})}\BibitemShut {NoStop}%
\bibitem [{\citenamefont {Erpenbeck}\ \emph
  {et~al.}(2023{\natexlab{a}})\citenamefont {Erpenbeck}, \citenamefont {Lin},
  \citenamefont {Blommel}, \citenamefont {Zhang}, \citenamefont {Iskakov},
  \citenamefont {Bernheimer}, \citenamefont {N\'u\~nez Fern\'andez},
  \citenamefont {Cohen}, \citenamefont {Parcollet}, \citenamefont {Waintal},\
  and\ \citenamefont {Gull}}]{Erpenbeck2023b}%
  \BibitemOpen
  \bibfield  {author} {\bibinfo {author} {\bibfnamefont {A.}~\bibnamefont
  {Erpenbeck}}, \bibinfo {author} {\bibfnamefont {W.-T.}\ \bibnamefont {Lin}},
  \bibinfo {author} {\bibfnamefont {T.}~\bibnamefont {Blommel}}, \bibinfo
  {author} {\bibfnamefont {L.}~\bibnamefont {Zhang}}, \bibinfo {author}
  {\bibfnamefont {S.}~\bibnamefont {Iskakov}}, \bibinfo {author} {\bibfnamefont
  {L.}~\bibnamefont {Bernheimer}}, \bibinfo {author} {\bibfnamefont
  {Y.}~\bibnamefont {N\'u\~nez Fern\'andez}}, \bibinfo {author} {\bibfnamefont
  {G.}~\bibnamefont {Cohen}}, \bibinfo {author} {\bibfnamefont
  {O.}~\bibnamefont {Parcollet}}, \bibinfo {author} {\bibfnamefont
  {X.}~\bibnamefont {Waintal}},\ and\ \bibinfo {author} {\bibfnamefont
  {E.}~\bibnamefont {Gull}},\ }\bibfield  {title} {\bibinfo {title} {Tensor
  train continuous time solver for quantum impurity models},\ }\href
  {https://doi.org/10.1103/PhysRevB.107.245135} {\bibfield  {journal} {\bibinfo
   {journal} {Phys. Rev. B}\ }\textbf {\bibinfo {volume} {107}},\ \bibinfo
  {pages} {245135} (\bibinfo {year} {2023}{\natexlab{a}})}\BibitemShut
  {NoStop}%
\bibitem [{\citenamefont {K\"unzel}\ \emph {et~al.}(2024)\citenamefont
  {K\"unzel}, \citenamefont {Erpenbeck}, \citenamefont {Werner}, \citenamefont
  {Arrigoni}, \citenamefont {Gull}, \citenamefont {Cohen},\ and\ \citenamefont
  {Eckstein}}]{Kunzel2024}%
  \BibitemOpen
  \bibfield  {author} {\bibinfo {author} {\bibfnamefont {F.}~\bibnamefont
  {K\"unzel}}, \bibinfo {author} {\bibfnamefont {A.}~\bibnamefont {Erpenbeck}},
  \bibinfo {author} {\bibfnamefont {D.}~\bibnamefont {Werner}}, \bibinfo
  {author} {\bibfnamefont {E.}~\bibnamefont {Arrigoni}}, \bibinfo {author}
  {\bibfnamefont {E.}~\bibnamefont {Gull}}, \bibinfo {author} {\bibfnamefont
  {G.}~\bibnamefont {Cohen}},\ and\ \bibinfo {author} {\bibfnamefont
  {M.}~\bibnamefont {Eckstein}},\ }\bibfield  {title} {\bibinfo {title}
  {Numerically exact simulation of photodoped {Mott} insulators},\ }\href
  {https://doi.org/10.1103/PhysRevLett.132.176501} {\bibfield  {journal}
  {\bibinfo  {journal} {Phys. Rev. Lett.}\ }\textbf {\bibinfo {volume} {132}},\
  \bibinfo {pages} {176501} (\bibinfo {year} {2024})}\BibitemShut {NoStop}%
\bibitem [{\citenamefont {Profumo}\ \emph {et~al.}(2015)\citenamefont
  {Profumo}, \citenamefont {Groth}, \citenamefont {Messio}, \citenamefont
  {Parcollet},\ and\ \citenamefont {Waintal}}]{Profumo2015}%
  \BibitemOpen
  \bibfield  {author} {\bibinfo {author} {\bibfnamefont {R.~E.~V.}\
  \bibnamefont {Profumo}}, \bibinfo {author} {\bibfnamefont {C.}~\bibnamefont
  {Groth}}, \bibinfo {author} {\bibfnamefont {L.}~\bibnamefont {Messio}},
  \bibinfo {author} {\bibfnamefont {O.}~\bibnamefont {Parcollet}},\ and\
  \bibinfo {author} {\bibfnamefont {X.}~\bibnamefont {Waintal}},\ }\bibfield
  {title} {\bibinfo {title} {Quantum {Monte} {Carlo} for correlated
  out-of-equilibrium nanoelectronic devices},\ }\href
  {https://doi.org/10.1103/PhysRevB.91.245154} {\bibfield  {journal} {\bibinfo
  {journal} {Phys. Rev. B}\ }\textbf {\bibinfo {volume} {91}},\ \bibinfo
  {pages} {245154} (\bibinfo {year} {2015})}\BibitemShut {NoStop}%
\bibitem [{\citenamefont {Moutenet}\ \emph {et~al.}(2019)\citenamefont
  {Moutenet}, \citenamefont {Seth}, \citenamefont {Ferrero},\ and\
  \citenamefont {Parcollet}}]{Moutenet2019}%
  \BibitemOpen
  \bibfield  {author} {\bibinfo {author} {\bibfnamefont {A.}~\bibnamefont
  {Moutenet}}, \bibinfo {author} {\bibfnamefont {P.}~\bibnamefont {Seth}},
  \bibinfo {author} {\bibfnamefont {M.}~\bibnamefont {Ferrero}},\ and\ \bibinfo
  {author} {\bibfnamefont {O.}~\bibnamefont {Parcollet}},\ }\bibfield  {title}
  {\bibinfo {title} {Cancellation of vacuum diagrams and the long-time limit in
  out-of-equilibrium diagrammatic quantum {Monte} {Carlo}},\ }\href
  {https://doi.org/10.1103/PhysRevB.100.085125} {\bibfield  {journal} {\bibinfo
   {journal} {Phys. Rev. B}\ }\textbf {\bibinfo {volume} {100}},\ \bibinfo
  {pages} {085125} (\bibinfo {year} {2019})}\BibitemShut {NoStop}%
\bibitem [{\citenamefont {Bertrand}\ \emph {et~al.}(2021)\citenamefont
  {Bertrand}, \citenamefont {Bauernfeind}, \citenamefont {Dumitrescu},
  \citenamefont {Ma\ifmmode~\check{c}\else \v{c}\fi{}ek}, \citenamefont
  {Waintal},\ and\ \citenamefont {Parcollet}}]{Bertrand2021}%
  \BibitemOpen
  \bibfield  {author} {\bibinfo {author} {\bibfnamefont {C.}~\bibnamefont
  {Bertrand}}, \bibinfo {author} {\bibfnamefont {D.}~\bibnamefont
  {Bauernfeind}}, \bibinfo {author} {\bibfnamefont {P.~T.}\ \bibnamefont
  {Dumitrescu}}, \bibinfo {author} {\bibfnamefont {M.}~\bibnamefont
  {Ma\ifmmode~\check{c}\else \v{c}\fi{}ek}}, \bibinfo {author} {\bibfnamefont
  {X.}~\bibnamefont {Waintal}},\ and\ \bibinfo {author} {\bibfnamefont
  {O.}~\bibnamefont {Parcollet}},\ }\bibfield  {title} {\bibinfo {title}
  {Quantum quasi {Monte} {Carlo} algorithm for out-of-equilibrium {Green}
  functions at long times},\ }\href
  {https://doi.org/10.1103/PhysRevB.103.155104} {\bibfield  {journal} {\bibinfo
   {journal} {Phys. Rev. B}\ }\textbf {\bibinfo {volume} {103}},\ \bibinfo
  {pages} {155104} (\bibinfo {year} {2021})}\BibitemShut {NoStop}%
\bibitem [{\citenamefont {Keiter}\ and\ \citenamefont
  {Kimball}(1970)}]{Keiter1971}%
  \BibitemOpen
  \bibfield  {author} {\bibinfo {author} {\bibfnamefont {H.}~\bibnamefont
  {Keiter}}\ and\ \bibinfo {author} {\bibfnamefont {J.~C.}\ \bibnamefont
  {Kimball}},\ }\bibfield  {title} {\bibinfo {title} {Perturbation technique
  for the {Anderson} {Hamiltonian}},\ }\href
  {https://doi.org/10.1103/PhysRevLett.25.672} {\bibfield  {journal} {\bibinfo
  {journal} {Phys. Rev. Lett.}\ }\textbf {\bibinfo {volume} {25}},\ \bibinfo
  {pages} {672} (\bibinfo {year} {1970})}\BibitemShut {NoStop}%
\bibitem [{\citenamefont {Bickers}(1987)}]{Bickers1987}%
  \BibitemOpen
  \bibfield  {author} {\bibinfo {author} {\bibfnamefont {N.~E.}\ \bibnamefont
  {Bickers}},\ }\bibfield  {title} {\bibinfo {title} {Review of techniques in
  the {large-$N$} expansion for dilute magnetic alloys},\ }\href
  {https://doi.org/10.1103/RevModPhys.59.845} {\bibfield  {journal} {\bibinfo
  {journal} {Rev. Mod. Phys.}\ }\textbf {\bibinfo {volume} {59}},\ \bibinfo
  {pages} {845} (\bibinfo {year} {1987})}\BibitemShut {NoStop}%
\bibitem [{\citenamefont {Coleman}(1984)}]{Coleman1984}%
  \BibitemOpen
  \bibfield  {author} {\bibinfo {author} {\bibfnamefont {P.}~\bibnamefont
  {Coleman}},\ }\bibfield  {title} {\bibinfo {title} {New approach to the
  mixed-valence problem},\ }\href {https://doi.org/10.1103/PhysRevB.29.3035}
  {\bibfield  {journal} {\bibinfo  {journal} {Phys. Rev. B}\ }\textbf {\bibinfo
  {volume} {29}},\ \bibinfo {pages} {3035} (\bibinfo {year}
  {1984})}\BibitemShut {NoStop}%
\bibitem [{\citenamefont {Werner}\ \emph {et~al.}(2006)\citenamefont {Werner},
  \citenamefont {Comanac}, \citenamefont {de' Medici}, \citenamefont {Troyer},\
  and\ \citenamefont {Millis}}]{Werner2006}%
  \BibitemOpen
  \bibfield  {author} {\bibinfo {author} {\bibfnamefont {P.}~\bibnamefont
  {Werner}}, \bibinfo {author} {\bibfnamefont {A.}~\bibnamefont {Comanac}},
  \bibinfo {author} {\bibfnamefont {L.}~\bibnamefont {de' Medici}}, \bibinfo
  {author} {\bibfnamefont {M.}~\bibnamefont {Troyer}},\ and\ \bibinfo {author}
  {\bibfnamefont {A.~J.}\ \bibnamefont {Millis}},\ }\bibfield  {title}
  {\bibinfo {title} {Continuous-time solver for quantum impurity models},\
  }\href {https://doi.org/10.1103/PhysRevLett.97.076405} {\bibfield  {journal}
  {\bibinfo  {journal} {Phys. Rev. Lett.}\ }\textbf {\bibinfo {volume} {97}},\
  \bibinfo {pages} {076405} (\bibinfo {year} {2006})}\BibitemShut {NoStop}%
\bibitem [{\citenamefont {Haule}(2024)}]{Haule2023}%
  \BibitemOpen
  \bibfield  {author} {\bibinfo {author} {\bibfnamefont {K.}~\bibnamefont
  {Haule}},\ }\href@noop {} {\bibinfo {title} {Strong coupling quantum impurity
  solver on the real and imaginary axis}} (\bibinfo {year} {2024}),\ \Eprint
  {https://arxiv.org/abs/2311.09412} {arXiv:2311.09412 [cond-mat.str-el]}
  \BibitemShut {NoStop}%
\bibitem [{\citenamefont {Eckstein}\ and\ \citenamefont
  {Werner}(2010)}]{Eckstein2010nca}%
  \BibitemOpen
  \bibfield  {author} {\bibinfo {author} {\bibfnamefont {M.}~\bibnamefont
  {Eckstein}}\ and\ \bibinfo {author} {\bibfnamefont {P.}~\bibnamefont
  {Werner}},\ }\bibfield  {title} {\bibinfo {title} {Nonequilibrium dynamical
  mean-field calculations based on the noncrossing approximation and its
  generalizations},\ }\href {https://doi.org/10.1103/PhysRevB.82.115115}
  {\bibfield  {journal} {\bibinfo  {journal} {Phys. Rev. B}\ }\textbf {\bibinfo
  {volume} {82}},\ \bibinfo {pages} {115115} (\bibinfo {year}
  {2010})}\BibitemShut {NoStop}%
\bibitem [{\citenamefont {Pruschke}\ and\ \citenamefont
  {Grewe}(1989)}]{Pruschke1989}%
  \BibitemOpen
  \bibfield  {author} {\bibinfo {author} {\bibfnamefont {T.}~\bibnamefont
  {Pruschke}}\ and\ \bibinfo {author} {\bibfnamefont {N.}~\bibnamefont
  {Grewe}},\ }\bibfield  {title} {\bibinfo {title} {The {Anderson} model with
  finite {Coulomb} repulsion},\ }\href@noop {} {\bibfield  {journal} {\bibinfo
  {journal} {Z. Phys. B Condens. Matter}\ }\textbf {\bibinfo {volume} {74}},\
  \bibinfo {pages} {439} (\bibinfo {year} {1989})}\BibitemShut {NoStop}%
\bibitem [{\citenamefont {Haule}\ \emph {et~al.}(2001)\citenamefont {Haule},
  \citenamefont {Kirchner}, \citenamefont {Kroha},\ and\ \citenamefont
  {W\"olfle}}]{Haule2013}%
  \BibitemOpen
  \bibfield  {author} {\bibinfo {author} {\bibfnamefont {K.}~\bibnamefont
  {Haule}}, \bibinfo {author} {\bibfnamefont {S.}~\bibnamefont {Kirchner}},
  \bibinfo {author} {\bibfnamefont {J.}~\bibnamefont {Kroha}},\ and\ \bibinfo
  {author} {\bibfnamefont {P.}~\bibnamefont {W\"olfle}},\ }\bibfield  {title}
  {\bibinfo {title} {{Anderson} impurity model at finite {Coulomb} interaction
  {U}: Generalized noncrossing approximation},\ }\href
  {https://doi.org/10.1103/PhysRevB.64.155111} {\bibfield  {journal} {\bibinfo
  {journal} {Phys. Rev. B}\ }\textbf {\bibinfo {volume} {64}},\ \bibinfo
  {pages} {155111} (\bibinfo {year} {2001})}\BibitemShut {NoStop}%
\bibitem [{\citenamefont {Oseledets}(2011)}]{Oseledets2011}%
  \BibitemOpen
  \bibfield  {author} {\bibinfo {author} {\bibfnamefont {I.~V.}\ \bibnamefont
  {Oseledets}},\ }\bibfield  {title} {\bibinfo {title} {Tensor-train
  decomposition},\ }\href {https://doi.org/10.1137/090752286} {\bibfield
  {journal} {\bibinfo  {journal} {SIAM Journal on Scientific Computing}\
  }\textbf {\bibinfo {volume} {33}},\ \bibinfo {pages} {2295} (\bibinfo {year}
  {2011})}\BibitemShut {NoStop}%
\bibitem [{\citenamefont {Dolgov}\ and\ \citenamefont
  {Savostyanov}(2020)}]{Dolgov2020}%
  \BibitemOpen
  \bibfield  {author} {\bibinfo {author} {\bibfnamefont {S.}~\bibnamefont
  {Dolgov}}\ and\ \bibinfo {author} {\bibfnamefont {D.}~\bibnamefont
  {Savostyanov}},\ }\bibfield  {title} {\bibinfo {title} {Parallel cross
  interpolation for high-precision calculation of high-dimensional integrals},\
  }\href {https://doi.org/https://doi.org/10.1016/j.cpc.2019.106869} {\bibfield
   {journal} {\bibinfo  {journal} {Computer Physics Communications}\ }\textbf
  {\bibinfo {volume} {246}},\ \bibinfo {pages} {106869} (\bibinfo {year}
  {2020})}\BibitemShut {NoStop}%
\bibitem [{\citenamefont {N\'u\~nez Fern\'andez}\ \emph
  {et~al.}(2022)\citenamefont {N\'u\~nez Fern\'andez}, \citenamefont {Jeannin},
  \citenamefont {Dumitrescu}, \citenamefont {Kloss}, \citenamefont {Kaye},
  \citenamefont {Parcollet},\ and\ \citenamefont {Waintal}}]{Fernandez2022}%
  \BibitemOpen
  \bibfield  {author} {\bibinfo {author} {\bibfnamefont {Y.}~\bibnamefont
  {N\'u\~nez Fern\'andez}}, \bibinfo {author} {\bibfnamefont {M.}~\bibnamefont
  {Jeannin}}, \bibinfo {author} {\bibfnamefont {P.~T.}\ \bibnamefont
  {Dumitrescu}}, \bibinfo {author} {\bibfnamefont {T.}~\bibnamefont {Kloss}},
  \bibinfo {author} {\bibfnamefont {J.}~\bibnamefont {Kaye}}, \bibinfo {author}
  {\bibfnamefont {O.}~\bibnamefont {Parcollet}},\ and\ \bibinfo {author}
  {\bibfnamefont {X.}~\bibnamefont {Waintal}},\ }\bibfield  {title} {\bibinfo
  {title} {Learning {Feynman} diagrams with tensor trains},\ }\href
  {https://doi.org/10.1103/PhysRevX.12.041018} {\bibfield  {journal} {\bibinfo
  {journal} {Phys. Rev. X}\ }\textbf {\bibinfo {volume} {12}},\ \bibinfo
  {pages} {041018} (\bibinfo {year} {2022})}\BibitemShut {NoStop}%
\bibitem [{\citenamefont {Fernández}\ \emph {et~al.}(2024)\citenamefont
  {Fernández}, \citenamefont {Ritter}, \citenamefont {Jeannin}, \citenamefont
  {Li}, \citenamefont {Kloss}, \citenamefont {Louvet}, \citenamefont
  {Terasaki}, \citenamefont {Parcollet}, \citenamefont {von Delft},
  \citenamefont {Shinaoka},\ and\ \citenamefont {Waintal}}]{Fernandez2024}%
  \BibitemOpen
  \bibfield  {author} {\bibinfo {author} {\bibfnamefont {Y.~N.}\ \bibnamefont
  {Fernández}}, \bibinfo {author} {\bibfnamefont {M.~K.}\ \bibnamefont
  {Ritter}}, \bibinfo {author} {\bibfnamefont {M.}~\bibnamefont {Jeannin}},
  \bibinfo {author} {\bibfnamefont {J.-W.}\ \bibnamefont {Li}}, \bibinfo
  {author} {\bibfnamefont {T.}~\bibnamefont {Kloss}}, \bibinfo {author}
  {\bibfnamefont {T.}~\bibnamefont {Louvet}}, \bibinfo {author} {\bibfnamefont
  {S.}~\bibnamefont {Terasaki}}, \bibinfo {author} {\bibfnamefont
  {O.}~\bibnamefont {Parcollet}}, \bibinfo {author} {\bibfnamefont
  {J.}~\bibnamefont {von Delft}}, \bibinfo {author} {\bibfnamefont
  {H.}~\bibnamefont {Shinaoka}},\ and\ \bibinfo {author} {\bibfnamefont
  {X.}~\bibnamefont {Waintal}},\ }\href {https://arxiv.org/abs/2407.02454}
  {\bibinfo {title} {Learning tensor networks with tensor cross interpolation:
  new algorithms and libraries}} (\bibinfo {year} {2024}),\ \Eprint
  {https://arxiv.org/abs/2407.02454} {arXiv:2407.02454 [physics.comp-ph]}
  \BibitemShut {NoStop}%
\bibitem [{\citenamefont {Erpenbeck}\ \emph
  {et~al.}(2023{\natexlab{b}})\citenamefont {Erpenbeck}, \citenamefont {Gull},\
  and\ \citenamefont {Cohen}}]{Erpenbeck2023}%
  \BibitemOpen
  \bibfield  {author} {\bibinfo {author} {\bibfnamefont {A.}~\bibnamefont
  {Erpenbeck}}, \bibinfo {author} {\bibfnamefont {E.}~\bibnamefont {Gull}},\
  and\ \bibinfo {author} {\bibfnamefont {G.}~\bibnamefont {Cohen}},\ }\bibfield
   {title} {\bibinfo {title} {Quantum {Monte} {Carlo} method in the steady
  state},\ }\href {https://doi.org/10.1103/PhysRevLett.130.186301} {\bibfield
  {journal} {\bibinfo  {journal} {Phys. Rev. Lett.}\ }\textbf {\bibinfo
  {volume} {130}},\ \bibinfo {pages} {186301} (\bibinfo {year}
  {2023}{\natexlab{b}})}\BibitemShut {NoStop}%
\bibitem [{\citenamefont {Kamenev}(2011)}]{KamenevBook}%
  \BibitemOpen
  \bibfield  {author} {\bibinfo {author} {\bibfnamefont {A.}~\bibnamefont
  {Kamenev}},\ }\href@noop {} {\emph {\bibinfo {title} {{Field Theory of
  Non-Equilibrium Systems}}}}\ (\bibinfo  {publisher} {Cambridge University
  Press},\ \bibinfo {address} {Cambridge, UK},\ \bibinfo {year}
  {2011})\BibitemShut {NoStop}%
\bibitem [{Note1()}]{Note1}%
  \BibitemOpen
  \bibinfo {note} {The strong coupling expansion can also be derived by
  introducing pseudo-particles for each many body state $|m\rangle $ on the
  impurity \cite {Coleman1984}. In this formulation, the resolvents $\protect
  \mathcal {G}$ have the meaning of ``pseudo-particles propagators'' and
  $\protect \mathcal {S}_{\protect \text {int}}$ corresponds to a
  ``pseudo-particle interaction''. We will adopt this terminology even though
  the derivation presented in App.~\protect \textup {\hbox {\mathsurround \z@
  \protect \normalfont (\ignorespaces \ref {app:strcpl}\unskip \@@italiccorr
  )}} does not introduce pseudo-particles.}\BibitemShut {Stop}%
\bibitem [{Note2()}]{Note2}%
  \BibitemOpen
  \bibinfo {note} {Symmetries imply a block-diagonal structure of these
  matrices, which is not indicated in the equations below, but used in the
  numerical implementation}\BibitemShut {NoStop}%
\bibitem [{\citenamefont {Sch\"uler}\ \emph {et~al.}(2020)\citenamefont
  {Sch\"uler}, \citenamefont {Golez}, \citenamefont {Murakami}, \citenamefont
  {Bittner}, \citenamefont {Herrmann}, \citenamefont {Strand}, \citenamefont
  {Werner},\ and\ \citenamefont {Eckstein}}]{Nessi}%
  \BibitemOpen
  \bibfield  {author} {\bibinfo {author} {\bibfnamefont {M.}~\bibnamefont
  {Sch\"uler}}, \bibinfo {author} {\bibfnamefont {D.}~\bibnamefont {Golez}},
  \bibinfo {author} {\bibfnamefont {Y.}~\bibnamefont {Murakami}}, \bibinfo
  {author} {\bibfnamefont {N.}~\bibnamefont {Bittner}}, \bibinfo {author}
  {\bibfnamefont {A.}~\bibnamefont {Herrmann}}, \bibinfo {author}
  {\bibfnamefont {H.~U.}\ \bibnamefont {Strand}}, \bibinfo {author}
  {\bibfnamefont {P.}~\bibnamefont {Werner}},\ and\ \bibinfo {author}
  {\bibfnamefont {M.}~\bibnamefont {Eckstein}},\ }\bibfield  {title} {\bibinfo
  {title} {Nessi: The non-equilibrium systems simulation package},\ }\href
  {https://doi.org/https://doi.org/10.1016/j.cpc.2020.107484} {\bibfield
  {journal} {\bibinfo  {journal} {Comput. Phys. Commun.}\ }\textbf {\bibinfo
  {volume} {257}},\ \bibinfo {pages} {107484} (\bibinfo {year}
  {2020})}\BibitemShut {NoStop}%
\bibitem [{\citenamefont {Golez}\ \emph {et~al.}(2024)\citenamefont {Golez},
  \citenamefont {Paprotzki}, \citenamefont {Werner},\ and\ \citenamefont
  {Eckstein}}]{Golez2024xas}%
  \BibitemOpen
  \bibfield  {author} {\bibinfo {author} {\bibfnamefont {D.}~\bibnamefont
  {Golez}}, \bibinfo {author} {\bibfnamefont {E.}~\bibnamefont {Paprotzki}},
  \bibinfo {author} {\bibfnamefont {P.}~\bibnamefont {Werner}},\ and\ \bibinfo
  {author} {\bibfnamefont {M.}~\bibnamefont {Eckstein}},\ }\href
  {https://arxiv.org/abs/2409.06314} {\bibinfo {title} {Measuring the ultrafast
  screening of {$U$} in photo-excited charge-transfer insulators with
  time-resolved {X-ray} absorption spectroscopy}} (\bibinfo {year} {2024}),\
  \Eprint {https://arxiv.org/abs/2409.06314} {arXiv:2409.06314
  [cond-mat.str-el]} \BibitemShut {NoStop}%
\bibitem [{\citenamefont {Werner}\ and\ \citenamefont
  {Millis}(2007)}]{Werner2007}%
  \BibitemOpen
  \bibfield  {author} {\bibinfo {author} {\bibfnamefont {P.}~\bibnamefont
  {Werner}}\ and\ \bibinfo {author} {\bibfnamefont {A.~J.}\ \bibnamefont
  {Millis}},\ }\bibfield  {title} {\bibinfo {title} {Efficient dynamical mean
  field simulation of the {Holstein}-{Hubbard} model},\ }\href
  {https://doi.org/10.1103/PhysRevLett.99.146404} {\bibfield  {journal}
  {\bibinfo  {journal} {Phys. Rev. Lett.}\ }\textbf {\bibinfo {volume} {99}},\
  \bibinfo {pages} {146404} (\bibinfo {year} {2007})}\BibitemShut {NoStop}%
\bibitem [{\citenamefont {Paprotzki}\ and\ \citenamefont
  {Eckstein}(2024)}]{Paprotzki2024}%
  \BibitemOpen
  \bibfield  {author} {\bibinfo {author} {\bibfnamefont {E.}~\bibnamefont
  {Paprotzki}}\ and\ \bibinfo {author} {\bibfnamefont {M.}~\bibnamefont
  {Eckstein}},\ }\href@noop {} {\bibinfo {title} {to be published}} (\bibinfo
  {year} {2024})\BibitemShut {NoStop}%
\bibitem [{\citenamefont {Strand}\ \emph {et~al.}(2024)\citenamefont {Strand},
  \citenamefont {Kleinhenz},\ and\ \citenamefont {Krivenko}}]{Strand2024}%
  \BibitemOpen
  \bibfield  {author} {\bibinfo {author} {\bibfnamefont {H.~U.~R.}\
  \bibnamefont {Strand}}, \bibinfo {author} {\bibfnamefont {J.}~\bibnamefont
  {Kleinhenz}},\ and\ \bibinfo {author} {\bibfnamefont {I.}~\bibnamefont
  {Krivenko}},\ }\bibfield  {title} {\bibinfo {title} {Inchworm quasi {Monte}
  {Carlo} for quantum impurities},\ }\href
  {https://doi.org/10.1103/PhysRevB.110.L121120} {\bibfield  {journal}
  {\bibinfo  {journal} {Phys. Rev. B}\ }\textbf {\bibinfo {volume} {110}},\
  \bibinfo {pages} {L121120} (\bibinfo {year} {2024})}\BibitemShut {NoStop}%
\bibitem [{\citenamefont {Ma\ifmmode~\check{c}\else \v{c}\fi{}ek}\ \emph
  {et~al.}(2020)\citenamefont {Ma\ifmmode~\check{c}\else \v{c}\fi{}ek},
  \citenamefont {Dumitrescu}, \citenamefont {Bertrand}, \citenamefont {Triggs},
  \citenamefont {Parcollet},\ and\ \citenamefont {Waintal}}]{Macek2020}%
  \BibitemOpen
  \bibfield  {author} {\bibinfo {author} {\bibfnamefont {M.}~\bibnamefont
  {Ma\ifmmode~\check{c}\else \v{c}\fi{}ek}}, \bibinfo {author} {\bibfnamefont
  {P.~T.}\ \bibnamefont {Dumitrescu}}, \bibinfo {author} {\bibfnamefont
  {C.}~\bibnamefont {Bertrand}}, \bibinfo {author} {\bibfnamefont
  {B.}~\bibnamefont {Triggs}}, \bibinfo {author} {\bibfnamefont
  {O.}~\bibnamefont {Parcollet}},\ and\ \bibinfo {author} {\bibfnamefont
  {X.}~\bibnamefont {Waintal}},\ }\bibfield  {title} {\bibinfo {title} {Quantum
  quasi-{Monte} {Carlo} technique for many-body perturbative expansions},\
  }\href {https://doi.org/10.1103/PhysRevLett.125.047702} {\bibfield  {journal}
  {\bibinfo  {journal} {Phys. Rev. Lett.}\ }\textbf {\bibinfo {volume} {125}},\
  \bibinfo {pages} {047702} (\bibinfo {year} {2020})}\BibitemShut {NoStop}%
\bibitem [{\citenamefont {Kim}\ \emph {et~al.}(2022)\citenamefont {Kim},
  \citenamefont {Li}, \citenamefont {Eckstein},\ and\ \citenamefont
  {Werner}}]{Kim2022}%
  \BibitemOpen
  \bibfield  {author} {\bibinfo {author} {\bibfnamefont {A.~J.}\ \bibnamefont
  {Kim}}, \bibinfo {author} {\bibfnamefont {J.}~\bibnamefont {Li}}, \bibinfo
  {author} {\bibfnamefont {M.}~\bibnamefont {Eckstein}},\ and\ \bibinfo
  {author} {\bibfnamefont {P.}~\bibnamefont {Werner}},\ }\bibfield  {title}
  {\bibinfo {title} {Pseudoparticle vertex solver for quantum impurity
  models},\ }\href {https://doi.org/10.1103/PhysRevB.106.085124} {\bibfield
  {journal} {\bibinfo  {journal} {Phys. Rev. B}\ }\textbf {\bibinfo {volume}
  {106}},\ \bibinfo {pages} {085124} (\bibinfo {year} {2022})}\BibitemShut
  {NoStop}%
\bibitem [{\citenamefont {Kim}\ \emph {et~al.}(2023)\citenamefont {Kim},
  \citenamefont {Lenk}, \citenamefont {Li}, \citenamefont {Werner},\ and\
  \citenamefont {Eckstein}}]{Kim2023}%
  \BibitemOpen
  \bibfield  {author} {\bibinfo {author} {\bibfnamefont {A.~J.}\ \bibnamefont
  {Kim}}, \bibinfo {author} {\bibfnamefont {K.}~\bibnamefont {Lenk}}, \bibinfo
  {author} {\bibfnamefont {J.}~\bibnamefont {Li}}, \bibinfo {author}
  {\bibfnamefont {P.}~\bibnamefont {Werner}},\ and\ \bibinfo {author}
  {\bibfnamefont {M.}~\bibnamefont {Eckstein}},\ }\bibfield  {title} {\bibinfo
  {title} {Vertex-based diagrammatic treatment of light-matter-coupled
  systems},\ }\href {https://doi.org/10.1103/PhysRevLett.130.036901} {\bibfield
   {journal} {\bibinfo  {journal} {Phys. Rev. Lett.}\ }\textbf {\bibinfo
  {volume} {130}},\ \bibinfo {pages} {036901} (\bibinfo {year}
  {2023})}\BibitemShut {NoStop}%
\bibitem [{\citenamefont {Kaye}\ \emph {et~al.}(2024)\citenamefont {Kaye},
  \citenamefont {Huang}, \citenamefont {Strand},\ and\ \citenamefont
  {Gole\ifmmode~\check{z}\else \v{z}\fi{}}}]{Kaye2024}%
  \BibitemOpen
  \bibfield  {author} {\bibinfo {author} {\bibfnamefont {J.}~\bibnamefont
  {Kaye}}, \bibinfo {author} {\bibfnamefont {Z.}~\bibnamefont {Huang}},
  \bibinfo {author} {\bibfnamefont {H.~U.~R.}\ \bibnamefont {Strand}},\ and\
  \bibinfo {author} {\bibfnamefont {D.}~\bibnamefont
  {Gole\ifmmode~\check{z}\else \v{z}\fi{}}},\ }\bibfield  {title} {\bibinfo
  {title} {Decomposing imaginary-time {Feynman} diagrams using separable basis
  functions: {Anderson} impurity model strong-coupling expansion},\ }\href
  {https://doi.org/10.1103/PhysRevX.14.031034} {\bibfield  {journal} {\bibinfo
  {journal} {Phys. Rev. X}\ }\textbf {\bibinfo {volume} {14}},\ \bibinfo
  {pages} {031034} (\bibinfo {year} {2024})}\BibitemShut {NoStop}%
\bibitem [{\citenamefont {Kaye}\ and\ \citenamefont
  {U.~R.~Strand}(2023)}]{Kaye2023b}%
  \BibitemOpen
  \bibfield  {author} {\bibinfo {author} {\bibfnamefont {J.}~\bibnamefont
  {Kaye}}\ and\ \bibinfo {author} {\bibfnamefont {H.}~\bibnamefont
  {U.~R.~Strand}},\ }\bibfield  {title} {\bibinfo {title} {A fast time domain
  solver for the equilibrium {Dyson} equation},\ }\href
  {https://doi.org/10.1007/s10444-023-10067-7} {\bibfield  {journal} {\bibinfo
  {journal} {Advances in Computational Mathematics}\ }\textbf {\bibinfo
  {volume} {49}},\ \bibinfo {pages} {63} (\bibinfo {year} {2023})}\BibitemShut
  {NoStop}%
\end{thebibliography}
%

\end{document}